\documentclass{article}

\usepackage{arxiv}

\usepackage{array}
\usepackage[caption=false,font=normalsize,labelfont=sf,textfont=sf]{subfig}
\usepackage{textcomp}
\usepackage{stfloats}
\usepackage{url}
\usepackage{verbatim}
\usepackage{graphicx}
\usepackage{cite}
\usepackage{amsmath,amssymb,amsfonts,amsthm}
\usepackage{algorithmic,algorithm}
\usepackage{textcomp}
\usepackage{xcolor}
\usepackage{subfig}
\usepackage{multirow}

\newtheorem{lemma}{Lemma}
\newtheorem{theorem}{Theorem}
\newtheorem{definition}{Definition}

\title{ZF Beamforming Tensor Compression for Massive MIMO Fronthaul}

\author{
 Libin Zheng \\
  School of Mathematics\\
  Hunan University\\
  Hunan, 410082 \\
  \texttt{fifholz301@hnu.edu.cn} \\
   \And
 Zihao Wang \\
 School of Mathematics\\
 Hunan University\\
 Hunan, 410082 \\
  \texttt{wzh97@hnu.edu.cn} \\
 \And
  Minru Bai \\
   School of Mathematics\\
   Hunan University\\
   Hunan, 410082 \\
 \texttt{minru-bai@hnu.edu.cn} \\
  \And
 Zhenjie Tan \\
 Huawei Technologies Co., Ltd.\\
 Sichuan 610036, China\\
  \texttt{tanzhenjie@huawei.com} \\
}

\begin{document}
\maketitle
\begin{abstract}
    In the rapidly evolving landscape of 5G and beyond 5G (B5G) mobile cellular communications, efficient data compression and reconstruction strategies become paramount, especially in massive multiple-input multiple-output (MIMO) systems. A critical challenge in these systems is the capacity-limited fronthaul, particularly in the context of the Ethernet-based common public radio interface (eCPRI) connecting baseband units (BBUs) and remote radio units (RRUs). This capacity limitation hinders the effective handling of increased traffic and data flows. We propose a novel two-stage compression approach to address this bottleneck. The first stage employs sparse Tucker decomposition, targeting the weight tensor's low-rank components for compression. The second stage further compresses these components using complex givens decomposition and run-length encoding, substantially improving the compression ratio. Our approach specifically targets the Zero-Forcing (ZF) beamforming weights in BBUs. By reconstructing these weights in RRUs, we significantly alleviate the burden on eCPRI traffic, enabling a higher number of concurrent streams in the radio access network (RAN). Through comprehensive evaluations, we demonstrate the superior effectiveness of our method in Channel State Information (CSI) compression, paving the way for more efficient 5G/B5G fronthaul links.

\textbf{Keywords:} MU-MIMO systems, Tensor compression, Downlink transmission, Tucker decomposition,  Fronthaul.
\end{abstract}

\section{Introduction}
Massive multi-user multiple-input and multiple-output (MU-MIMO) systems are pivotal in advancing the capabilities of fifth-generation (5G) and beyond 5G (B5G) mobile cellular communications. These systems enable base stations, equipped with multiple antennas, to simultaneously communicate with various user equipments. By leveraging spatial diversity and multiplexing techniques, MU-MIMO systems facilitate higher data rates and more reliable communication links, achieved through the strategic deployment of pre-designed beamforming weights.

The architecture of modern 5G/B5G base stations (BS) consists of two primary components: baseband units (BBUs) and remote radio units (RRUs). BBUs are responsible for high-level baseband functions, including modulation of user data into specific quadrature amplitude modulation (QAM) signals and the calculation of transmit beamforming weights based on predefined criteria. RRUs, on the other hand, handle radio functions, including the application of these beamforming weights to user signals, thereby creating targeted beams. These units are interconnected via an optical-fiber-based fronthaul interface, typically the Ethernet-based common public radio interface (eCPRI), a standardized solution in the wireless industry. However, as the demand for wireless services grows, the capacity limitations of the eCPRI fronthaul interface become increasingly apparent. This limitation poses significant challenges for downlink transmissions in massive MU-MIMO systems, where the traffic of user data and beamforming weights is increasing. Therefore, innovative solutions are essential to alleviate fronthaul traffic while preserving the efficiency of MU-MIMO systems. A critical aspect of this challenge is finding effective methods to compress beamforming weights, which currently consume a substantial portion of eCPRI capacity.

In MU-MIMO systems, classical beamforming techniques like transmit Wiener filtering (WF), zero-forcing (ZF), and maximum ratio transmission (MRT) play a crucial role in computing the transmit beamforming matrix \cite{b8, b9}. ZF precoding, in particular, stands out for its simplicity and low computational complexity. This method does not demand knowledge of channel statistics or SNR, and it can be efficiently implemented using matrix inversion techniques. A notable advantage of ZF precoding is its capacity to completely eliminate multi-user interference, thereby achieving optimal Signal-to-Interference-plus-Noise Ratio (SINR) performance. However, as the number of users, antennas, and subcarriers increase, the computational and storage costs of the ZF weights rise, so it needs to be compressed.

Traditional approaches, such as compressive sensing\cite{CS, donoho2009message, daubechies2004iterative}, have been employed to minimize feedback data in these systems. These methods typically utilize the inherent sparsity of data, especially after transformations like the Discrete Fourier Transform (DFT). However, compressive sensing relies heavily on the assumption that the data exhibits a high degree of sparsity. Our experiments indicate that this assumption does not strongly hold for ZF-weighted data. In recent times, deep learning (DL) based methods have gained popularity in the context of channel state information (CSI) compression \cite{DL1, DL2, CSINN1}. These learning-based methods come with substantial training costs. Additionally, it's not always straightforward to adapt methods designed for CSI compression to the compression of beamforming weights, presenting a challenge in their direct application. For channels, there has been work on estimating and compressing channels using the low-rank property of CSI\cite{lrce, lrcc}. However, the large size of CSI data makes it difficult to compute the weights on the RRU. As for the weight tensor, its low-rank property is not so significant. For the weights tensor, on the other hand, the low rank is not significant. In summary, because the weight tensor lacks strong sparsity and low rank, compressing it with low error is a major challenge.

In our study, we adopt tensors as a high-order data format to represent Zero-Forcing (ZF) weights across the antenna, frequency, and user domains. Tensors offer a more comprehensive framework for data representation in MIMO-related problems, surpassing the conventional use of vectors and matrices (first-order and second-order tensors, respectively). 

Tensor technology can be widely used in MIMO applications\cite{tbprecode2021, beamtrain2021, 9739774}, especially in channel estimation\cite{cpdCE, CE1, CE2, CE3, CE4, CE5, DLCE}. \cite{TDinMimo} provided an in-depth description of tensor-decomposition-based methods for blind symbol recovery and channel parameter estimation, focusing on CDMA, STF, cooperative/relay, MIMO, and mmWave systems. Among various tensor decompositions, CANDECOMP/PARAFAC decomposition (CPD) \cite{b10} and Tucker decomposition (TD) \cite{b11} are widely recognized. Within the area of wireless communications, \cite{cpdCE} model the training signals in MIMO-OFDM systems assisted by intelligent reconfigurable surfaces (IRSs) as CPD tensor with missing fibers or slices by leveraging the sparse characteristics of high-frequency propagation, and estimate MS-to-IRS and BS-to-IRS channels by tensor completion. Tucker decomposition has been effectively utilized in 3D-MIMO systems for the rotation matrix in rotated codebooks, addressing the challenges of high dimensionality, as highlighted in prior studies\cite{b5,b7}.

The efficacy of Tucker decomposition in compressing tensor data, particularly when the data displays sparsity in its core tensor with a low Tucker rank, is well-documented. The Tucker decomposition is widely used to compress high-dimensional data such as high-resolution images\cite{b6}, hyperspectral images\cite{li2021correlation, li2020correlation} and neural network\cite{yin2020compressing, liu2022deep, cai2020learning}. In previous research, following Tucker decomposition, truncation and quantization techniques were applied to the core tensor for compression\cite{b12, ballard2020tuckermpi, beaupere2023higher}. However, there are often a large number of tiny elements in the core tensor produced by decomposition, and truncating them directly produces large errors. An alternative approach, TTHRESH\cite{b6}, merges high order singular value decomposition (HOSVD) with run-length coding, showing promise in compressing high-resolution visual images with favorable compression ratios. Nevertheless, the main compression space of TTHRESH lies in the number of bits, while the weight tensor uses fewer bits, and TTHRESH is based on HOSVD, which does not utilize the low-rank nature that the weight tensor has well, and thus the compression effect is not good at the weight tensor. At the same time, these methods mentioned above are basically designed for real tensors, not for complex tensors, and their quantization strategies are not necessarily optimal. Conventional weight compression methods\cite{R15, R16} usually use the discrete Fourier transform (DCT) to decompose the channel matrices in both the null and frequency domains, transform them to an energy-concentrated domain, and then quantize them in this transform domain. However the DCT basis vectors are not necessarily an optimal decomposition basis for compression. In our numerical experiments, each of these methods had difficulty in maintaining low rate loss at very low compression rates (e.g., 10\% or 15\%), as can be seen in the tables \ref{tab:eMBB} following the text. All of the above compression methods face great challenges in compressing the weights tensor at a low rate loss, so how to design the compression method for the data characteristics of the weights tensor becomes a key issue.

To overcome this challenge, we propose a two-stage compression approach. The first stage utilizes the low-rank nature of the weight tensor to try to solve a low-rank tucker decomposition approximating the original tensor, while ensuring that the core tensor is as sparse as possible. This is equivalent to solving a transformation for each dimension of the weight tensor, which allows the transformed tensor to be sparse and easy to quantize. In the second stage we quantize the sparse transformation tensor as well as the semi-orthogonal transformations, where for the sparse tensor we use sparse coding, and for the transformation matrix we perform a complex givens decomposition followed by RLE.
In this study, our objective is to efficiently compress the ZF weight tensor for massive MU-MIMO systems, thereby optimizing the utilization of downlink fronthaul bandwidth, particularly in Ethernet-based common public radio interface (eCPRI) environments. 
Our approach ensures that only compressed beamforming weights and user data are delivered via the downlink fronthaul, with decompression occurring at the remote radio units (RRUs). This scheme significantly alleviates bandwidth constraints in the eCPRI for massive MU-MIMO systems. The key contributions of our work are outlined as follows:

\begin{itemize}
    \item \textbf{Proposing a two-stage compression method:} We propose a two-stage compression approach customized for ZF weight data, where the first stage obtains the sparse tensor and semi-orthogonal transformations by solving the sparse Tucker decomposition, while the second stage utilizes their data properties for quantization and compression.
    \item \textbf{Development of sparse tucker decomposition (STD) models for compression tensor with corresponding algorithms and convergence analysis:} We introduce a sparse tucker decomposition model, which helps us to find an optimal transformation of the weight tensor along each direction so that we can make the weight tensor as sparse as possible in the transformation domain, thus achieving better compression. We present a fast algorithm designed for solving STD models and prove its convergence as an algorithm with acceleration.
    \item \textbf{Factor Compression with Innovative Encoding and Quantization Strategy:} We devise a encoding and quantization strategy for the sparse core tensor and factor matrices, incorporating complex Givens decomposition and run-length encoding. This strategy is carefully calibrated to ensure optimal compression performance while minimally impacting the utility of the precoding.
    \item \textbf{Experimental Validation:} Through extensive numerical simulations, we validate the efficacy of our proposed method. Our results clearly demonstrate its capability to compress the weight tensor effectively. For the weight tensor with 128 antennas of transmitter, 136 resource blocks, 8 users equipments and 2 streams, we can achieve about 11\% compression ratio with only 5\% rate loss.
\end{itemize}

The structure of this paper is designed to methodically present our research and findings. Section 2 lays the groundwork by formulating the problem, focusing on the specific challenges associated with ZF weight tensor compression in massive MU-MIMO systems. Section 3 introduces our compression framework, detailing the decomposition model and the factor compression method. In Section 4, we describe the algorithm developed to solve our model, highlighting its design and optimization. Section 5 presents the numerical results, demonstrating the effectiveness and practicality of our approach. The paper concludes with Section 6, where we summarize our key findings and discuss the implications and potential future directions of our work.

\subsection{Preliminaries}
Now we give some notations, as show in Table \ref{tab:notations}. 

\begin{table}[]
    \centering
    \caption{Notations}\label{tab:notations}
    \begin{tabular}{l|l}
    \hline
    \multicolumn{1}{c|}{Variables}                                      & \multicolumn{1}{c}{Definitions}            \\ \hline
    Lower case letters: $x,y$                         & Vector                 \\ \hline
    Upper case bold letters: $\mathbf{X}, \mathbf{Y}$ & Matrix                 \\ \hline
    Calligraphic letters: $\mathcal{X}, \mathcal{Y}$  & Tensor                 \\ \hline
    \multicolumn{1}{c|}{$\mathbb{C}$}                                      & Set of complex numbers              \\ \hline
    \multicolumn{1}{c|}{$\mathbf{I}$}                                      & Identity matrix              \\ \hline
    \multicolumn{1}{c|}{$|\cdot|$ }                                     & Absolute value              \\ \hline
    \multicolumn{1}{c|}{$\|\cdot\|_F$}                                      & Frobenius norm              \\ \hline
    \multicolumn{1}{c|}{$\|\cdot\|_0$}                                      & $L_0$ norm              \\ \hline
    \multicolumn{1}{c|}{$(\cdot)^{\mathrm{H}}$}                                      & Conjugate transpose              \\ \hline
    \end{tabular}
\end{table}

Giving an $m$ th-order tensor denoted as $\mathcal{V} \in \mathbb{C}^{n_1\times \ldots \times n_m}$, the mode-$i$ product is denoted as $\mathcal{V}\times_i\mathbf{A}$, where $1\leq i\leq m$ and $\mathbf{A}\in\mathbb{C}^{n_i\times l}$. The size of resultant tensor is $n_1\times n_2 \times \cdots \times n_{i-1}\times l \times n_{i+1} \times \cdots \times n_{m-1} \times n_m$ and element wise product is $\left(\mathcal{V}\times_i\mathbf{A}\right)_{j_1\cdots j_{i-1}kj_{i+1}\cdots j_m}=\sum_{j_i=1}^{n_i}\mathcal{V}_{j_1j_2\cdots j_m}\mathbf{A}_{j_ik}$.

\begin{figure}[h]
    \centering
    \includegraphics[width=0.8\linewidth]{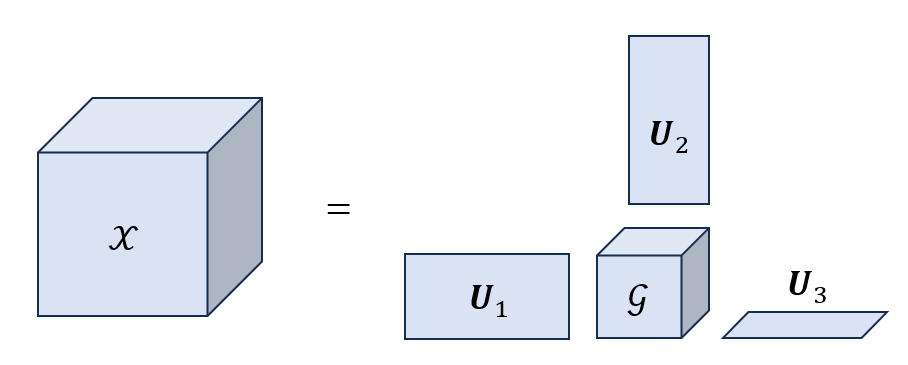}
    \caption{Tucker Decomposition}\label{fig:TD}
\end{figure}

The Tucker decomposition is a method of decomposing a tensor into a product of a core tensor and factor matrices in each direction, as show in Fig. \ref{fig:TD}. It can be written as $\mathcal{V}= \mathcal{G}\times_1\mathbf{U}_1 \times_2 \cdots \times_m \mathbf{U}_m$, where we will usually require the factor matrix to be column orthogonal. If we consider the factor matrices as some kind of column orthogonal transformation, then we can say that $\mathcal{G}$ is the tensor obtained by the corresponding transformations of $\mathcal{V}$ along each direction, i.e., $\mathcal{G}$ is the transformed tensor of $\mathcal{V}$ obtained by $\mathbf{U}_1, \mathbf{U}_2, \cdots \mathbf{U}_m$-transformations. In the following content, we will use $[\![\mathcal{G}; \mathbf{U}_1 , \cdots , \mathbf{U}_m]\!]$ to present Tucker decomposition for short. 

\section{Problem Formulation}

\begin{figure*}[!t]
    \centering
    \includegraphics[width=0.8\linewidth]{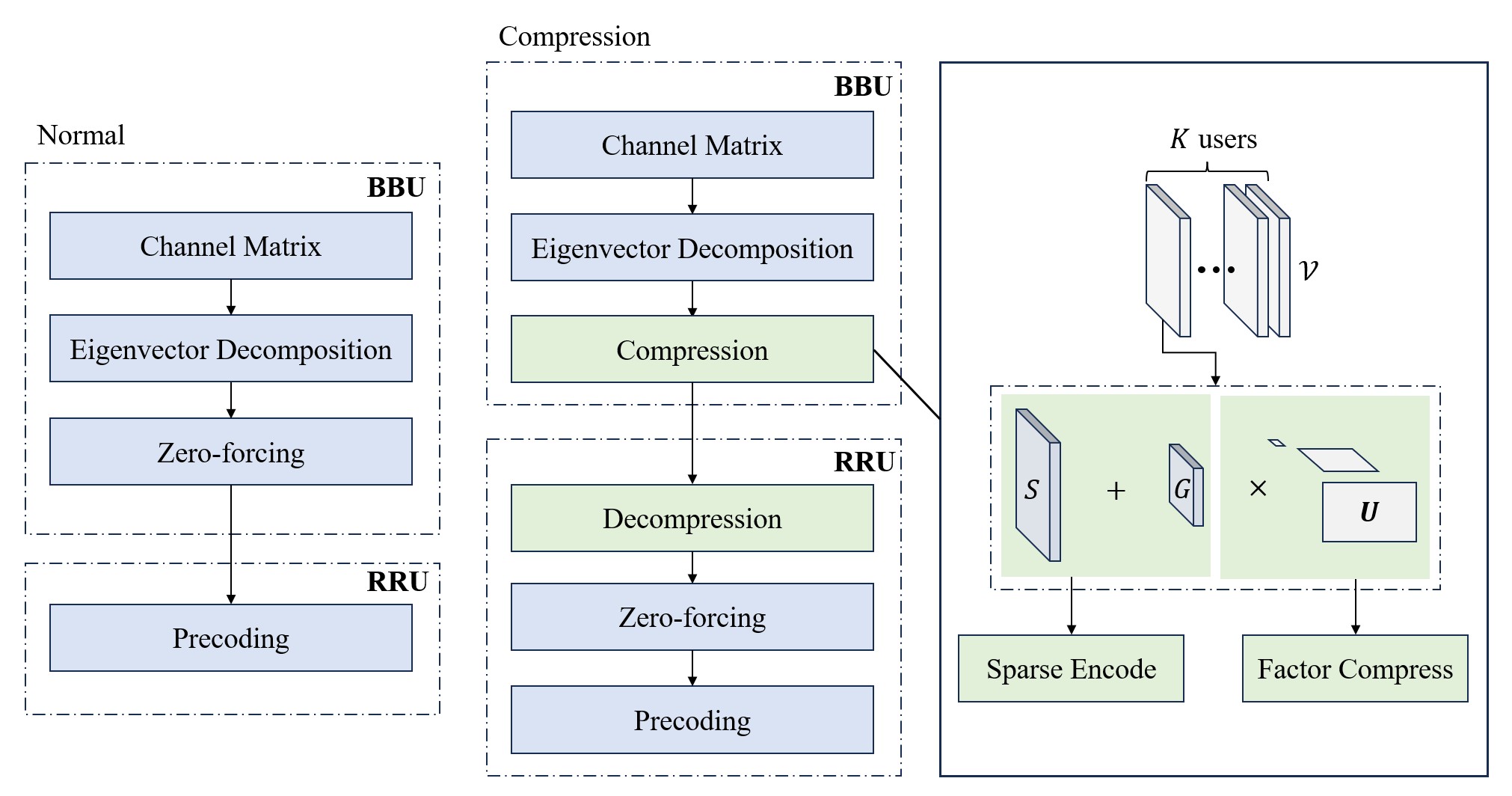}
    \caption{Compression framework}\label{fig:process}
\end{figure*}

In this study, we focus on the downlink of a multi-user MIMO (MU-MIMO) system. Consider a scenario with \( K \) users, \( J \) subcarriers, a base station (BS) equipped with \( N_t \) antennas, and each user equipped with \( N_u \) antennas, resulting in a total of \( KN_u \) antennas across all users. The received signal \( \mathbf{y}_k[j] \) for user \( k \) at \( j \) is given by:
\[
\mathbf{y}_k[j] = \sum_{i=1}^K \mathbf{H}_k[j] \mathbf{x}_i[j] + \mathbf{n}_k[j],
\]
where \( \mathbf{H}_k[j] \in \mathbb{C}^{N_{u} \times N_t} \) represents the channel matrix from the BS to user \( k \), \( \mathbf{x}_k[j] \in \mathbb{C}^{M \times 1} \) denotes the transmitted data matrix for user \( k \), and \( \mathbf{n}_k[j] \in \mathbb{C}^{N_{r,k}} \) is the additive complex Gaussian noise.

We employ Eigen-based Zero-forcing (ZF) beamforming, which offers a favorable balance between complexity and performance. Assuming data transmission across \( r \) parallel streams, the ZF beamforming process can be described as:
\[
\mathbf{W}_k[j]=\mathbf{V}_k[j]^H\left(\mathbf{V}_k[j] \mathbf{V}_k[j]^H\right)^{-1},
\]
where \( \mathbf{V}_k[j] \) is the eigenvector matrix corresponding to the \( r \) largest eigenvalues of \( \mathbf{H}_k[j]^H\mathbf{H}_k[j] \). These matrices \( {\mathbf{V}}_{k}[j] \) are combined into a third-order tensor \( \mathcal{V}_k \in \mathbb{C}^{r\times N_t\times J} \), representing the number of streams, antennas at the BS, and frequency elements, respectively.

For computational efficiency, \( \mathbf{V} \) is typically computed for a resource block (RB), which includes several resource elements (REs). In this case, \( \mathbf{V}_k[i] \) corresponds to the eigenvectors of the aggregated largest eigenvalues of \( \mathbf{H}_k[j]^H\mathbf{H}_k[j] \) for each RE \( j \) within RB \( i \). For simplicity, our analysis assumes \( 1 \) RB equals \( 1 \) RE, though in numerical experiments, we use \( 1 \) RB comprising \( 12 \) or \( 24 \) REs. Our problem is to compress the $\mathcal{V}_k$. 

\section{Two Stage Compression Approach}
Next we give a two-stage compression method for the above problem. The first stage obtains the sparse tensor and semi-orthogonal transformations by solving the sparse Tucker decomposition, while the second stage utilizes their data properties for quantization and compression. As illustrated in Figure \ref{fig:process}, our compression strategy involves calculating \( \mathcal{V} \) from CSI data at the BBU, compressing it, and then transmitting this compressed version via the eCPRI fronthaul. This approach eliminates the need to transmit the beamforming weight data directly. At the RRU, we decompress the \( \mathcal{V} \) data and subsequently compute \( \mathcal{W} \). 

\subsection{Sparse Tucker Decomposition (STD) Model}
In this section, we introduce the Sparse Tucker Decomposition (STD) model for efficient data compression. Our model decomposes the tensor into a sum of a sparse tensor and a low-rank tensor, where the low-rank tensor is represented as a product of a sparse core tensor and column-orthogonal factor matrices. The framework of our compression method is shown in Fig. \ref{fig:process}.

\begin{figure*}[htbp]
    \centering
    \subfloat[]{\includegraphics[width=0.3\linewidth]{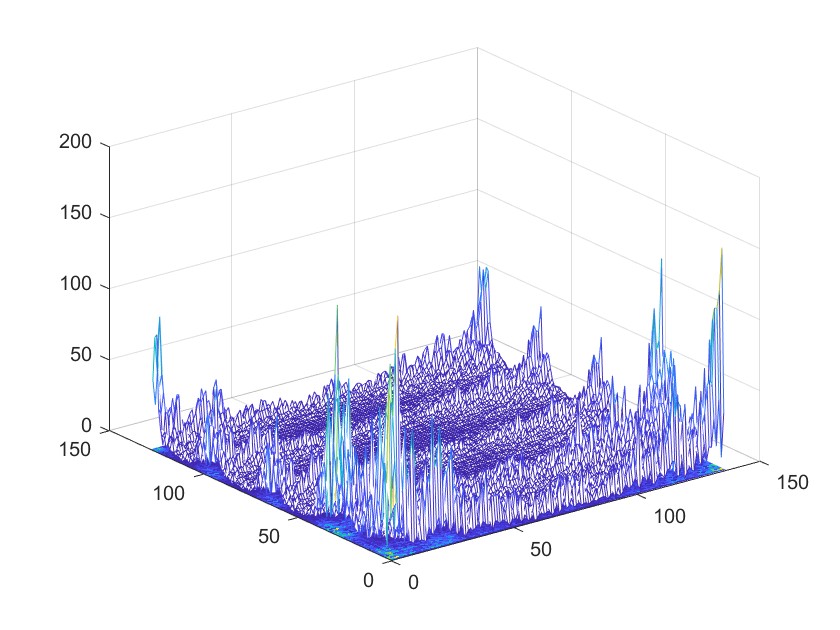}\label{fig:trans_V}}
    \subfloat[]{\includegraphics[width=0.3\linewidth]{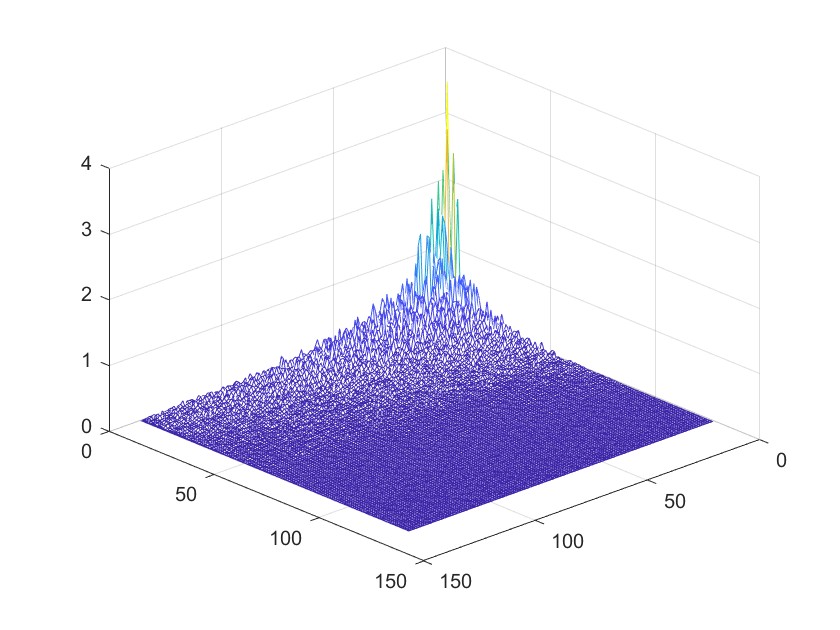}\label{fig:TDcore}}
    \subfloat[]{\includegraphics[width=0.3\linewidth]{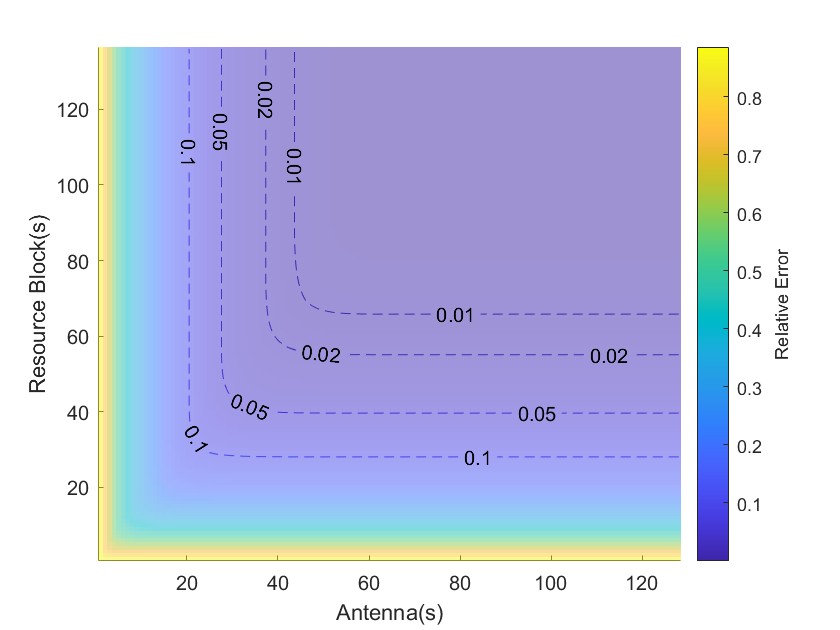}\label{fig:lowrank}}
    \caption{(a) Weight Matrix under DFT. (b) Core tensor under Tucker Decomposition. (c) Relative errors of the tensor produced by tucker decomposition for different tucker ranks and the original tensor, the dashed lines are contours for relative errors of 0.01, 0.02, 0.05 and 0.1, respectively.}
\end{figure*}

\textbf{Why Use a Low-Rank Tensor to Approximate Tensor \( \mathcal{V}_k \)?} Experiments reveal that the tensor \( \mathcal{V}_k \) for each user is approximately low-rank. For instance, a \( 2 \times 128 \times 136 \) tensor can be approximated within a 5\% relative error using a Tucker rank of \( 2 \times 30 \times 40 \). More details could be found in Fig \ref{fig:lowrank}.

\textbf{Why Decompose a Sparse Tensor from the To-Be-Compressed Tensor?} After applying a discrete Fourier transform to the BS antenna and frequency dimensions of tensor \( \mathcal{V} \), we observed strong energy concentration at the frequency extremes, suggesting that significant information can be efficiently represented as a sparse tensor.

\textbf{Why Use Sparse Tucker Decomposition?} High order singular value decomposition (HOSVD) often results in a core tensor with many small elements, as show in \ref{fig:TDcore}, which are difficult to truncate without loss of accuracy. However, it reveals such a fact: the core tensor $\mathcal{G}$ can be sparse under a particular transformation $\mathbf{U}_1, \mathbf{U}_2$ and $\mathbf{U}_3$. 

Based on the above analysis, we decide to compress $\mathcal{V}$ by decomposing $\mathcal{V}$ after DFT into the sum of tensor $\mathcal{L}$ and tensor $\mathcal{S}$, where tensor $\mathcal{S}$ is sparse on the original domain that has not been $\mathbf{U}$-transformed, while $\mathcal{L}$ is sparse on the $\mathbf{U}$-transformed domain. We will utilize sparse coding to compress these two sparse tensors.

\textbf{Sparse Encoding.} In our setting, for the sparse tensor $\mathcal{S}\in\mathbb{C}^{n_1\times n_2\times n_3\times n_4}$, we will split it into a sequential vector of non-zero elements and a binary tensor that records the positions where the elements exist. Thus we end up with the number of bits we need to store as the number of bits needed to store the non-zero digits plus the number of bits needed to store the binary coding tensor, which is of size $log_2(n_1\times n_2\times n_3\times n_4)$.

Our experiments demonstrate that decomposing the tensor \( \mathcal{V} \) into a low-rank tensor and a sparse tensor significantly improves its low-rank property, thus enhancing compression. We propose an STD model that employs Tucker decomposition to capture the low-rank structure, and the \( L_0 \) norm to encapsulate the sparsity of the tensor. Additionally, we introduce sparse constraints on the Tucker decomposition's core tensor for better compression of low-rank terms. The STD model is formulated as:
\begin{align}
    \nonumber \min_{\mathcal{G},\mathbf{U}_1,\mathbf{U}_2,\mathbf{U}_3,\mathcal{S}} & \frac{1}{2} \| \mathcal{V}-\mathcal{S}- [\![\mathcal{G}; \mathbf{U}_1 ,\mathbf{U}_2 ,\mathbf{U}_3 ]\!]\|_F^2\\
    \label{model}\text{s.t. } & 
   \mathbf{U}_i^H\mathbf{U}_i=\mathbf{I}, i=1,\cdots,3\\
    \nonumber& \|\mathcal{G}\|_0 \leq \alpha_1,\ \|\mathcal{S}\|_0 \leq \alpha_2,
\end{align}
where \( \mathcal{V}\in\mathbb{C}^{n_1\times n_2\times n_3} \) is the weight tensor for one user, \( \mathcal{S}\in\mathbb{C}^{n_1\times n_2\times n_3} \) is the sparse tensor; \( \mathcal{G}\in\mathbb{C}^{r_1\times r_2\times r_3} \) is the core tensor; \( \mathbf{U}_1\in\mathbb{C}^{r_1\times n_1}, \mathbf{U}_2\in\mathbb{C}^{r_2\times n_2}, \mathbf{U}_3\in\mathbb{C}^{r_3\times n_3} \) are factor matrices; and \( \alpha_1, \alpha_2 \) are the user-difined sparsity parameters that can be used to control the sparsity of \(\mathcal{G}, \mathcal{S}\) by the following equation
\begin{equation}\label{spar}
    \alpha_1=\lfloor s_1r_1r_2r_3\rfloor, \alpha_2=\lfloor s_2n_1n_2n_3\rfloor,
\end{equation}
where $s_1, s_2$ denote the sparsity of \(\mathcal{G}, \mathcal{S}\). The Tucker rank is denoted as \( [r_1, r_2, r_3] \). \(r_1, r_2, r_3, s_1\) and \(s_2\) help us to initially control our final compression rate so that it won't be higher than the \( \frac{s_1r_1r_2r_3+s_2n_1n_2n_3+r_1n_1+r_2n_2+r_3n_3}{n_1n_2n_3} \).

\subsection{Factor Compress}
In STD, a significant portion of the storage requirement is attributed to the factor matrices. For instance, in the case of a \(2 \times 128 \times 136\) weights tensor approximated by a \(2 \times 30 \times 40\) Tucker rank, the factors constitute approximately 26\% of the overall elements in the Tucker decomposition.

Inspiring by \cite{b6}, we try to further compress our cores and factors on bit-plane, where our main compression target is the factor. However, the method given in the \cite{b6} cannot be directly used in our model because of two differences: first, \cite{b6} targets the real tensor, while we are a complex tensor; second, because of the differences in the data size and decomposition method, our factor matrix is larger than the one in the article, and instead the storage occupies a higher percentage. Therefore we need to design specialized methods to compress the factors.

\begin{figure*}
    \centering
    \includegraphics[width=0.8\linewidth]{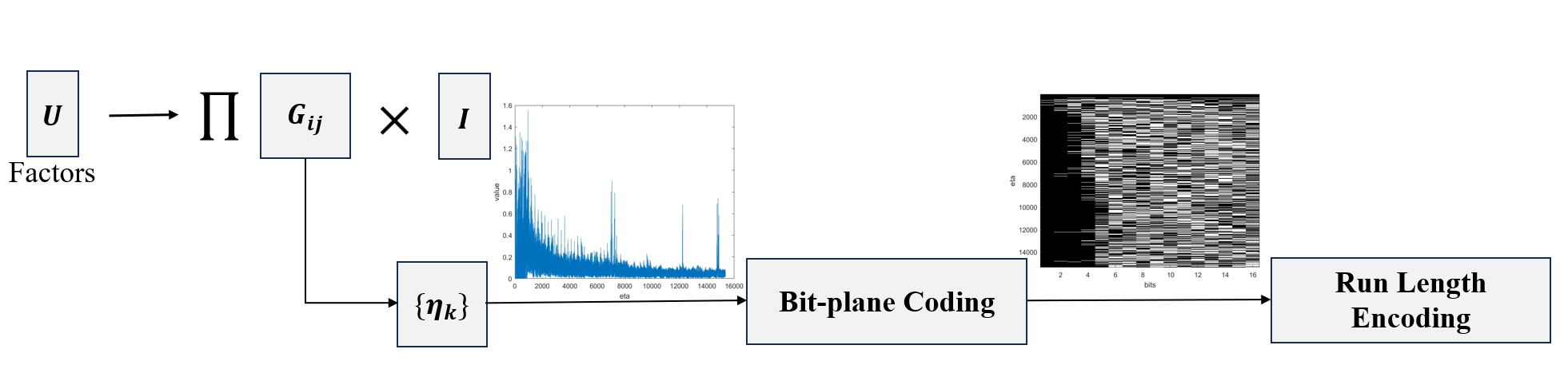}
    \caption{Factor Compression Framework}\label{fig:factorcompress}
\end{figure*}

Our entire compression process of the factor matrices is depicted in Fig. \ref{fig:factorcompress}. To enhance the compression of the tensor, we firstly adopt a strategy of parameterizing the factor matrices using Givens matrices\cite{b4}. This approach leverages the column orthogonality of the factor matrices. Subsequent analysis reveals that the parameter sequences derived from this method exhibit partial sparsity. By exploiting this sparsity, we can further compress these parameters using bit-plane coding\cite{b6}, thus achieving more efficient storage.

For the factor matrix \( U \in \mathbb{C}^{n \times r} \) (where \( n \geq r \)), full storage is not required due to the orthogonality of its columns. Instead, \( U \) can be decomposed into a product of \( \frac{(2n-r-1)r}{2} \) orthogonal matrices:
\[
U = \left(\prod_{i=1}^{r}\prod_{j=i+1}^{n}G_{ij}\right)\bar{I},
\]
where \( G_{ij} \in \mathbb{C}^{n \times n} \) is a sequence of orthogonal matrices for parameterization, and \( \bar{I} \in \mathbb{C}^{n \times r} \) is a diagonal matrix with 1s on the diagonal.

Givens transformations are utilized for these orthogonal matrices:
\[
G_{ij}([i,j],[i,j]) = \left(\begin{array}{cc}
\cos (\eta_k) & e^{j \theta_k} \sin (\eta_k) \\
-e^{-j \theta_k} \sin (\eta_k) & \cos (\eta_k)
\end{array}\right),
\]
with \( k = \frac{(2n-i)(i-1)}{2} + j - i \). Each Givens matrix requires only two real parameters, \( \eta_k \in [0,\pi] \) and \( \theta_k \in [0,2\pi] \), allowing the entire unitary matrix to be expressed using \( (2n-r-1)r \) real parameters.

Upon representing the factor matrix in this manner, we observe that the \( {\eta_k} \) sequence exhibits a degree of sparsity. While there are few absolute zeros, the sequence is dominated by smaller elements and sparse larger ones. The complete algorithmic procedure is shown in Algorithm \ref{algorithm:fc}, where the details of the Encode function can be found in the \cite{b6}.

\begin{algorithm}
    \caption{Obtain and Compress the Decomposition Parts of STD}    \label{algorithm:fc}
      \begin{algorithmic}[1]
          \item[1] {\bf Input}: $\mathcal{V}$ and $s_\mathcal{G}$
          \item[2] Compute $\mathcal{G}, \mathbf{U}_1, \mathbf{U}_2, \mathbf{U}_3$ and $\mathcal{S}$ by algorithm \ref{algorithm BPL1}.
          \item[3] $s_\mathbf{U}=Encode(\mathcal{G},s_\mathcal{G})$
          \item[4] for i = 1:3
          \item[5] $\quad[\eta_i, \theta_i]=ComplexGivens(\mathbf{U}_i)$
          \item[6] $\quad Encode(\mathbf{U}_i, s_\mathbf{U})$
          \item[7] end for 
    \end{algorithmic}
\end{algorithm}

\section{Accelerated Block Coordinate Descent Algorithm for STD}
In this section, we will propose an algorithm to solve the model (\ref{model}). We can rewrite model (\ref{model}) in the following form,
\begin{equation}
    \begin{aligned}
        & \min_{\mathcal{X}}\  F\left(\mathcal{X}\right):=\frac{1}{2} \| \mathcal{V}-\mathcal{S}- [\![\mathcal{G}; \mathbf{U}_1 ,\mathbf{U}_2 ,\mathbf{U}_3 ]\!]\|_F^2\\
       \label{model1} &+\delta_{\|\mathcal{G}\|_0 \leq \alpha_1}(\mathcal{G})
        +\delta_{\|\mathcal{S}\|_0 \leq \alpha_2}(\mathcal{S})+\sum_{i=1}^3\delta_{\mathbf{U}_i^H\mathbf{U}_i=\mathbf{I}}(\mathbf{U}_i)
    \end{aligned}
\end{equation}
where $\mathcal{X}=(\mathcal{G},\mathbf{U}_1,\mathbf{U}_2,\mathbf{U}_3,\mathcal{S})$ and $\delta_{\mathbb{C}}(\cdot)$ represent the  indicator function of the set $\mathbb{C}$. We adopt accelerated proximal block coordinate descent (APBCD) framework to design the algorithm. Denote $\mathbf{\bar{U}}_i$ is the factor matrice after acceleration, the algorithm solves the following subproblems.

\subsection{Algorithm for STD model}
First we will use the proximal operator. The proximal operator of a function $F\left(x\right)$ is defined as \[prox_{\lambda f}(x)=\mathop{\text{argmin}}_{y}\left\{\lambda F\left(x\right)+\frac{1}{2}\|x-y\|_F^2\right\}.\] The proximal operator of $l_0$ norm can be given by \[\left[prox_{\lambda \|\cdot\|_0}(x)\right]_i=\\ \left\{\begin{array}{cl}
    x_i, & \left|x_i\right|>\sqrt{2\alpha} \\
    0, & \left|x_i\right|\leq\sqrt{2\alpha}
    \end{array}   \right.   .\]

For core tensor $\mathcal{G}$, we have
    \begin{equation}\label{ox1}
        \begin{aligned}
            \mathcal{G}^{k+1}=&\mathop{\text{argmin}}_{\mathcal{G}}\left\{\delta_{\|\mathcal{G}\|_0 \leq \alpha_1}(\mathcal{G})+\frac{1}{2}\|\mathcal{G}-\mathcal{G}^k\|_F^2+\frac{\eta_\mathcal{G}^k}{2} \|\mathcal{V}-\mathcal{S}^k- [\![\mathcal{G}; \mathbf{\bar{U}}_1^k ,\mathbf{\bar{U}}_2^k ,\mathbf{\bar{U}}_3^k ]\!]\|_F^2\right\}\\
            =&\mathop{\text{argmin}}_{\mathcal{G}}\left\{\delta_{\|\mathcal{G}\|_0 \leq \alpha_1}+\frac{1}{2}\|\mathcal{G}-\mathcal{A}^k\|\right\}\\
            =&\mathop{\text{argmin}}_{\mathcal{G}}\left\{\lambda_{\mathcal{A}}^{\alpha_1}\|\mathcal{G}\|_0+\frac{1}{2}\|\mathcal{G}-\mathcal{A}^k\|\right\} \\
            =&prox_{\lambda_{\mathcal{A}}^{\alpha_1} \|\cdot\|_0}(\mathcal{A}^k),
        \end{aligned}
    \end{equation}
    where \(\mathcal{A}^k=\frac{\eta_\mathcal{G}^k [\![\mathcal{V}-\mathcal{S}^k; \mathbf{U}_1^{k^H} ,\mathbf{U}_2^{k^H} ,\mathbf{U}_3^{k^H} ]\!]+\mathcal{G}^k}{\eta_\mathcal{G}^k+1}\) and \(\lambda_{\mathcal{A}}^{\alpha_1}\) is the $\lfloor\alpha_1\rfloor$-th largest number of $\mathcal{A}$. 

For factor matrices $\mathbf{U}_1,\mathbf{U}_2,\mathbf{U}_3$, we have
    \begin{equation}\label{U}
        \begin{aligned}
            &\mathbf{U}_i^{k+1} =\mathop{\text{argmin}}_{\mathbf{\bar{U}}_i}\left\{\delta_{\mathbf{U}_i^H\mathbf{U}_i=\mathbf{I}}(\mathbf{U}_i)+ \frac{1}{2}\|\mathbf{U}_i-\mathbf{\bar{U}}_i^k\|_F^2+\frac{\eta_i^k}{2} \|\mathcal{V}-\mathcal{S}^k-[\![\mathcal{G}^k;\{\mathbf{\bar{U}}^{k+1}_{j< i}\}, \mathbf{U}_i ,\{\mathbf{\bar{U}}^{k}_{j> i}\} ]\!]\|_F^2\right\},
        \end{aligned}
    \end{equation}
    which has following closed-form solution\cite{U},
        \begin{equation}
        \begin{aligned}
            \label{oz1}&\mathbf{U}_i^{k+1} ={\mathbf{W}}_i\mathbf{Q}_i^H,
        \end{aligned}
    \end{equation}
        where 
        \begin{equation*}
            \begin{aligned}
                &[{\mathbf{W}}_i,-,\mathbf{Q}_i]=svd((\mathcal{V}-\mathcal{S})_{(i)}[\mathcal{G}^k;\{\mathbf{U}^{k+1}_{j< i}\}, \mathbf{U}_i ,\{\mathbf{U}^{k}_{j> i}\} ]\!]_{(i)}+\frac{1}{\eta_i^k}\mathbf{\bar{U}}_i^k).
            \end{aligned}
        \end{equation*}

For sparse tensor $\mathcal{S}$, we have
        \begin{equation}\label{S}
        \begin{aligned}
            &\mathcal{S}^{k+1}= \mathop{\text{argmin}}_{\mathcal{S}} \left\{\delta_{\|\mathcal{S}\|_0 \leq \alpha_2}(\mathcal{S})+\frac{1}{2}\|\mathcal{S}-\mathcal{S}^k\|_F^2+ \frac{\eta_\mathcal{S}^k}{2} \|\mathcal{V}-\mathcal{S}- [\![\mathcal{G}^{k+1}; \mathbf{\bar{U}}_1^{k+1} ,\mathbf{\bar{U}}_2^{k+1} ,\mathbf{\bar{U}}_3^{k+1} ]\!]\|_F^2\right\},
        \end{aligned}
    \end{equation}
    which has following closed-form solution,
    \begin{equation}
        \begin{aligned}
            \label{oy1}
            \mathcal{S}^{k+1}=  prox_{\lambda_{\mathcal{B}}^{\alpha_2} \|\cdot\|_0}(\mathcal{B}),                  
        \end{aligned}
    \end{equation}where $\mathcal{B}=\frac{\eta_\mathcal{S}^k\mathcal{V}-\eta_\mathcal{S}^k[\![\mathcal{G}^{k+1}; \mathbf{\bar{U}}_1^{k+1} ,\mathbf{\bar{U}}_2^{k+1} ,\mathbf{\bar{U}}_3^{k+1} ]\!]+\mathcal{S}^k}{\eta_\mathcal{S}^k+1}.$

Then we propose an accelerated block coordinate descent algorithm for STD model and it is summarized in Algorithm \ref{algorithm BPL1}. Since STD Model is a non-convex model, we use the inertial method. 

\begin{algorithm}
    \caption{Accelerated proximal block coordinate descent algorithm (APBCD) for STD}    \label{algorithm BPL1}
          \begin{algorithmic}[1]
              \item[1] {\bf Input}: $\mathcal{V}$.
              \item[2] {\bf Initialize}: $\mathcal{G}^0,\mathbf{U}_1^0,\mathbf{U}_2^0,\mathbf{U}_3^0,\mathcal{S}^0$. Set $k=0$, $t_0=1$.
              \item[3] Update $\mathcal{G}^{k+1}$ by  (\ref{ox1});   
    \item[4] for i=1:3
    \item[5] $\quad$Update  $\mathbf{U}_i^{k+1}$ by (\ref{oz1}).
     \item[5]$\quad$Update $\bar{\mathbf{U}}_i^{k+1}$ by $\bar{\mathbf{U}}_i^{k+1}=\mathbf{U}_i^{k+1}+\beta(\mathbf{U}_i^{k+1}-\bar{\mathbf{U}}_i^{k})$.
    
    \item[6] end for
    \item[7] Update $\mathcal{S}^{k+1}$ by  (\ref{oy1}); 
    
    \item[8] If converge, set $\mathcal{G}^{\ast}=\mathcal{G}^{k+1}$, $\mathbf{U}_1^{\ast}:=\mathbf{U}_1^{k+1}$, $\mathbf{U}_2^{\ast}:=\mathbf{U}_2^{k+1}$,
    $\mathbf{U}_3^{\ast}:=\mathbf{U}_3^{k+1}$ and
    $\mathcal{S}^{\ast}=\mathcal{S}^{k+1}$; else, set $k=k+1$, return to 2;
    \item[9]{\bf Output}: $ \mathcal{G}^{\ast}$, $ \mathbf{U}_1^{\ast}$, $ \mathbf{U}_2^{\ast}$, $ \mathbf{U}_3^{\ast}$, $\mathcal{S}^{\ast}$.
        \end{algorithmic}
\end{algorithm}

\subsection{Convergence Analysis of Algorithm \ref{algorithm BPL1}}
Now we establish the global convergence of the APBCD algorithm. Because of the acceleration, there is no way to guarantee that $F\left(x\right)$ is decreasing in every iteration, so we need to introduce an auxiliary function $H$, with the help of which we get the convergence of the algorithm. Denote that $\mathcal{W}=(\mathcal{G},\mathbf{U}_1,\mathbf{U}_2,\mathbf{U}_3,\mathcal{S},\mathbf{\bar{U}}_1,\mathbf{\bar{U}}_2,\mathbf{\bar{U}}_3)$, Lemma \ref{lemma:H1} will show that the function value of $H$ decreases sufficiently during the iterations, Lemma \ref{lemma:H2} will show that the subgradient of $H$ at $\left\{\mathcal{W}^k\right\}$ is bounded, and they will help to derive that the clustered points of $\left\{\mathcal{W}^k\right\}$ is contained in the set of critical points of $H$, as shown in Lemma \ref{lemma:crit}. Recall the notation $\mathcal{X}=(\mathcal{G},\mathbf{U}_1,\mathbf{U}_2,\mathbf{U}_3,\mathcal{S})$, eventually we will use these Lemma to derive our convergence Theorem that the $\mathcal{X}^k$ converges to the critical points of $F$ under specific assumptions. The relevant proofs can be found in the appendix.

We will first give the construction of $H$ and its descent lemma. 

% for simplify, we define a fuzhu function, X->W, H(X^k) decresing

\begin{lemma}\label{lemma:H1}
    Assume the sequence generated by Algorithm \ref{algorithm BPL1} which is denoted as $\{\mathcal{X}^{k}\}_{k\in\mathbb{N}}$ is bounded. There exists $\eta_{\mathcal{G}}>0, \eta_{\mathcal{S}}>0, \eta_i>0, \gamma_i>0, i=1,2,3$ and $\rho_1>0$ such that the following assertions hold.
     \begin{enumerate}
        \item Denote $u^k_i=\left\|\mathbf{U}_i^k-\bar{\mathbf{U}}_i^k\right\|_F^2+\left\|\mathbf{U}_i^{k+1}-\bar{\mathbf{U}}_i^{k+1}\right\|_F^2$, we have \begin{equation}
            \begin{aligned}
                \rho_1\left[\sum_{i=1}^3u^k_i+\|\mathcal{S}^{k+1}-\mathcal{S}^k\|_F^2+\|\mathcal{G}^{k+1}-\mathcal{G}^k\|_F^2\right] H\left(\mathcal{W}^k\right)-H\left(\mathcal{W}^{k+1}\right)&,
            \end{aligned}
        \end{equation} where $H$ is an auxiliary function defined as\[H(\mathcal{W})=F\left(\mathcal{X}\right)+\sum_{i=1}^3\gamma_i\|\mathbf{U}_i-\mathbf{\bar{U}}_i\|_F^2.\]
        \item $ \sum_{k=0}^{+\infty}\left\|\mathcal{X}^{k+1}-\mathcal{X}^k\right\|^2_F<+\infty, $ which also implies $\left\|\mathcal{X}^{k+1}-\mathcal{X}^k\right\|_F\rightarrow 0$ and $\left\|\bar{\mathbf{U}}^{k}-\mathbf{U}^k\right\|_F\rightarrow 0$ when $k\rightarrow+\infty$.
     \end{enumerate}

        % \mathbf{U}_i^k-\bar{\mathbf{U}}_i^k\right\|^2<+\infty,\ \sum_{k=0}^{+\infty}\left\|\mathcal{G}^{k+1}-\mathcal{G}^k\right\|^2<+\infty,\ \sum_{k=0}^{+\infty}\left\|\mathcal{S}^{k+1}-\mathcal{S}^k
\end{lemma}

In order to show that the algorithm converges to the critical point, we will study the subgradient of the auxiliary function $H$ at the iteration point. In the following Lemma we will show that the subgradient of $H$ at the iteration point is bounded.

\begin{lemma}\label{lemma:H2}
    Assume the sequence generated by Algorithm \ref{algorithm BPL1} which is denoted as $\{\mathcal{X}^{k}\}_{k\in\mathbb{N}}$ is bounded. Suppose $\eta_{\mathcal{G}}, \eta_{\mathcal{S}}$ and $H(\mathcal{W})$ are set as Lemma \ref{lemma:H1}. For all $k\geq 0$, define $$\begin{aligned}
        &C^{k+1}=\\
        &(C_{\mathcal{G}}^{k+1}, C_{11}^{k+1}, C_{12}^{k+1}, C_{13}^{k+1}, C_{\mathcal{S}}^{k+1}, C_{21}^{k+1}, C_{22}^{k+1}, C_{23}^{k+1}),
    \end{aligned}$$ where $$
    \left\{\begin{aligned}
    C_{\mathcal{G}}^{k+1}=&\nabla_{\mathcal{G}}H(\mathcal{G}^{k+1},\mathbf{U}_1^{k+1},\mathbf{U}_2^{k+1},\mathbf{U}_3^{k+1},\mathcal{S}^{k+1})\\
    &-\nabla_{\mathcal{G}}H(\mathcal{G}^{k+1},\mathbf{\bar{U}}_1^k,\mathbf{\bar{U}}_2^k,\mathbf{\bar{U}}_3^k,\mathcal{S}^{k})-\frac{1}{\eta_{\mathcal{G}}^k}(\mathcal{G}^{k+1}-\mathcal{G}^k) \\
    C_{1i}^{k+1}=
    & \nabla_{\mathbf{U}_i}H(\mathcal{G}^{k+1},\mathbf{U}_1^{k+1},\mathbf{U}_2^{k+1},\mathbf{U}_3^{k+1},\mathcal{S}^{k+1}) \\
    &-\nabla_{\mathbf{U}_i}H(\mathcal{G}^{k+1},\{\mathbf{\bar{U}}^{k+1}_{j< i}\}, \mathbf{U}^{k+1}_i ,\{\mathbf{\bar{U}}^{k}_{j> i}\},\mathcal{S}^{k})-\frac{1}{\eta_{i}^k}(\mathbf{U}^{k+1}-\mathbf{\bar{U}}^k)+2 \gamma_i(\mathbf{U}^{k+1}_i-\mathbf{\bar{U}}^{k+1}_i)\\
    C_{\mathcal{S}}^{k+1}=
    &\nabla_{\mathcal{S}}H(\mathcal{G}^{k+1},\mathbf{U}_1^{k+1},\mathbf{U}_2^{k+1},\mathbf{U}_3^{k+1},\mathcal{S}^{k+1})\\
    &-\nabla_{\mathcal{S}}H(\mathcal{G}^{k+1},\mathbf{\bar{U}}_1^{k+1},\mathbf{\bar{U}}_2^{k+1},\mathbf{\bar{U}}_3^{k+1},\mathcal{S}^{k+1})-\frac{1}{\eta_{\mathcal{S}}^k}(\mathcal{S}^{k+1}-\mathcal{S}^k) \\
    C_{2i}^{k+1}=
    &2 \gamma_i(\mathbf{\bar{U}}^{k+1}_i-\mathbf{U}^{k+1}_i) .
    \end{aligned}\right.
    $$
    Then $C^{k+1}\in\partial H(\mathcal{W}^{k+1})$, and there exist $\rho_2>0$ such that $$\begin{aligned}
        \frac{1}{\rho_2}\|C^{k+1}\|_F\leq&\|\mathcal{G}^{k+1}-\mathcal{G}^k\|_F+\|\mathcal{S}^{k+1}-\mathcal{S}^k\|_F +\sum_{i=1}^3\|\mathbf{\bar{U}}_i^{k+1}-\mathbf{U}^{k+1}_i\|_F.
    \end{aligned}$$
\end{lemma}

Combining the Lamma \ref{lemma:H1} and Lamma \ref{lemma:H2}, we can prove the following lamma. % result

\begin{lemma}\label{lemma:crit} % to therom
    Assume the sequence generated by Algorithm \ref{algorithm BPL1} which is denoted as $\{\mathcal{X}^{k}\}_{k\in\mathbb{N}}$ is bounded. Suppose $\eta_{\mathcal{G}}, \eta_{\mathcal{S}}$ and $H(\mathcal{W})$ are set as Lemma \ref{lemma:H1}. Then the following assertions hold.
    \begin{enumerate}
        \item The cluster point set of the sequence $\{\mathcal{W}^k\}$ which is denoted by $\Omega^*$ is nonempty compact set, and 
        \begin{equation}
            \lim_{k\rightarrow+\infty}\textrm{dist}(\mathcal{W}^k,\Omega^*)=0.
        \end{equation}
        \item Every cluster point of $\left\{\mathcal{W}^k\right\}$ is a critical point of $H$ and the function $H$ is constant on $\Omega^*$.
        \item If \(\mathcal{W}^c=(\mathcal{G}^c,\mathbf{U}^c_1,\mathbf{U}^c_2,\mathbf{U}^c_3,\mathcal{S}^c,\mathbf{\bar{U}}^c_1,\mathbf{\bar{U}}^c_2,\mathbf{\bar{U}}^c_3)\) is a cluster point of $\left\{\mathcal{W}^k\right\}$, then \(\mathcal{X}^c=(\mathcal{G}^c,\mathbf{U}^c_1,\mathbf{U}^c_2,\mathbf{U}^c_3,\mathcal{S}^c)\) is a critical point of the function $F$.
    \end{enumerate}
\end{lemma}

Lemma \ref{lemma:crit} gives that the sequence convergence point is critical point, further, using Lemma \ref{lemma:H1}, Lemma \ref{lemma:H2} and the KL property we can prove that the whole sequence is a Cauchy sequence and thus obtain that the whole sequence converges to the critical point, which is our main result Theorem \ref{theorem:converge}. Detailed proofs of the above lemmas are given in appendix.

\begin{theorem}\label{theorem:converge}
    Assume the sequence generated by Algorithm \ref{algorithm BPL1} which is denoted as $\{\mathcal{X}^{k}\}_{k\in\mathbb{N}}$ is bounded. Suppose $\eta_{\mathcal{G}}, \eta_{\mathcal{S}}$ and $H(\mathcal{W})$ are set as Lemma \ref{lemma:H1}. Then $$\sum_{k=0}^{+\infty}\|\mathcal{X}^{k+1}-\mathcal{X}^k\|_F<+\infty,$$ and  $\{\mathcal{X}^k\}$ converges to a critical point of $F$. 
\end{theorem}

\section{Numerical Results}
This section explores the application of our proposed Hybrid Sparse Tucker Decomposition (STD) model, as detailed in (\ref{model}), for channel data compression in a downlink massive MIMO system setup.

\subsection{Evaluation indicators and Simulation Settings}
We consider a system comprising \( K \) users, each communicating over \( J \) resource blocks (RBs) with a base station equipped with \( N_t \) antennas. Every user \( k \) employs a signal \( \mathbf{s}_k[j] \in \mathbb{C}^{L} \) in a given RB \( j \), utilizing a precoding matrix \( \bar{\mathbf{W}}_k[j] \in \mathbb{C}^{N_t \times r} \). The channel matrix from the BS to user \( k \) is denoted as \( \mathbf{H}_k[j] \in \mathbb{C}^{N_{u} \times N_t} \). The signal received by user \( k \) comprises transmissions from all users, expressed as:
\[
\mathbf{y}_k[j] = \sum_{i=1}^K \mathbf{H}_k[j] \bar{\mathbf{W}}_i[j] \mathbf{s}_i[j] + \mathbf{n}_k[j].
\]

Our study particularly focuses on enhanced Mobile Broadband (eMBB) scenarios. We conduct numerical experiments to assess the efficacy of the STD method in compressing weights, benchmarking it against existing methods like  Tucker truncation (TD)\cite{b12}, TTHRESH\cite{b6} and R16 ETypeII\cite{R16}. The experiments utilize two channel datasets derived from the TR 38.901 protocol: dataset 1 involves \( N_t = 128 \) BS antennas and \( J = 136 \) RBs; dataset 2 consists of \( N_t = 512 \) BS antennas and \( J = 544 \) RBs. Both datasets feature \( K = 8 \) user equipments (UEs) and \( L = 2 \) data streams, with 16-bit precoding tensor data in the eCPRI link. The dataset 1 channel represents the current state of 5G technology, exhibiting a relatively small scale and less pronounced tensor low-rankness, thus posing challenges in achieving high compression without affecting the rate requirements. In contrast, the dataset 2 channel, reflective of future technology trends with broader carrier ranges and larger BS antenna arrays, shows higher low-rank characteristics and compressibility, suggesting potential for more advanced technology implementations at this scale. % the sparsity of G.

We will use the sum rate as the performance metric. The data rate of stream $l$, RB $j$ can be written as \cite{b13}
\begin{align*}
\mathrm{C}_{l,j}&=n_{k_l}[j] \mathbf{1}+
\sum_{i \neq l}  \mathbf{H}_{k_l}[j] \bar{\mathbf{W}}_i[j] \bar{\mathbf{W}}_i[j]^H \mathbf{H}_{k_l}[j]^H,\\
\mathrm{R}_{l,j}&=\log_2\left|1+\bar{\mathbf{W}}_l[j]^H \mathbf{H}_{k_l}[j]^H\mathrm{C}_{l,j}^{-1}\mathbf{H}_{k_l}[j] \bar{\mathbf{W}}_l[j]
\right|,
\end{align*}
and the sum rate is $\sum_{l}^{KL}\sum_{j}^{J}\mathrm{R}_{l,j}$, where ${k_l}$ denotes the user of stream $l$ and $\bar{\mathbf{W}}[j]$ denotes the compressed weight values vector of stream $l$ in RB $j$. The compression ratio $(CR)$ is defined as the number of transmitted bits required for the compressed data divided by the number of transmitted bits required for the weights $\mathcal{V}$. $\mathcal{G}$ sparsity and $\mathcal{S}$ sparsity refers to the percentage of the number of nonzero elements in the tensor, and is related to $\alpha_1$ and $\alpha_2$ as described by the equation (\ref{spar}).

\begin{table}[]
    \centering
    \caption{Simulation parameters}
    \begin{tabular}{lc}
    \hline \multicolumn{2}{c}{Abbreviate} \\ \hline
    Relative Error & RE \\     
    Rate Loss & RL \\ 
    Compression Rate & CR \\
    \hline \multicolumn{2}{c}{dataset 1} \\ \hline
    Resource blocks \(J\) & 136 \\     
    Number \( N_t \) of antennas at transmitter & 128 \\ 
    Users \( K \) & 8         \\
    Streams \( r \) & 2        \\ 
    SNR (dB) & 20        \\    
    Quantized bits & 16 \\
    \hline \multicolumn{2}{c}{dataset 2} \\ \hline
    Resource blocks \(J\) & 544 \\  
    Number \( N_t \) of antennas at transmitter & 512 \\  
    Users \( K \) & 8         \\
    Streams \( r \) & 2        \\ 
    SNR (dB) & 20        \\ 
    Quantized bits & 16 \\ \hline
    \end{tabular}
\end{table}

\subsection{Convergence}
\begin{figure}
    \centering
    \includegraphics[width=0.5\linewidth]{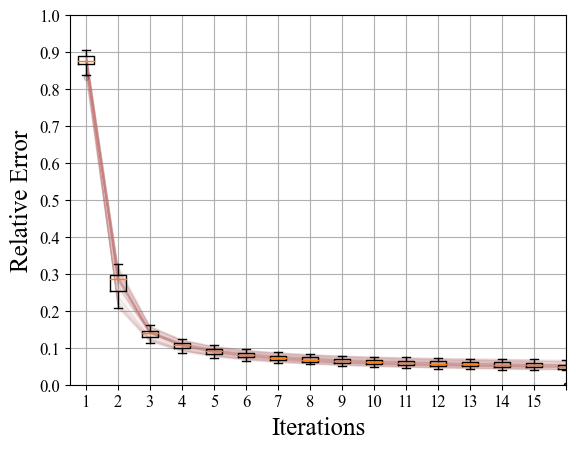}
    \caption{Plot of relative error between compressed and original weights with iteration steps for 300 generated weight datas.}\label{fig:iters}
\end{figure}

First, we will experimentally demonstrate the convergence of our algorithm. We generate 300 weights by protocol, compress them, and observe their relative errors to the original weights. The relevant results are shown in Fig. \ref{fig:iters}. Here we use a Tucker rank of $2\times 30\times 40$ and a core sparsity of 50\%, plus an $\mathcal{S}$-tensor of 1\% sparsity. It can be seen that our algorithm can quickly converge to a desired value in about 10 steps. Further iterations later on the rate of decline will gradually weaken, and eventually the whole curve flattens out. 

\subsection{Model Validity}
\begin{table}[]
    \centering
    \caption{The compression effect of different methods on the protocol generation channel, the lowest rate loss that can be achieved by each method when satisfying $CR\leq 10\%$ and $CR\leq 15\%$ respectively.}\label{tab:eMBB}
    \begin{tabular}{|c|rr|rr|}
    \hline
    \multicolumn{1}{|l|}{\multirow{2}{*}{Method}} & \multicolumn{2}{c|}{dataset 1}                        & \multicolumn{2}{c|}{dataset 2}                                  \\ \cline{2-5} 
    \multicolumn{1}{|l|}{}           & \multicolumn{1}{c}{CR} & \multicolumn{1}{c|}{RL} & \multicolumn{1}{c}{CR}      & \multicolumn{1}{c|}{RL}      \\ \hline\hline
    \multirow{2}{*}{TTHRESH\cite{b6}}         & 9.13\%                 & 93.97\%                 & 9.78\%                      & 94.28\%                      \\
                                     & 14.14\%                & 86.74\%                 & 14.74\%                     & 91.31\%                      \\ \hline
    \multirow{2}{*}{R16-eTypeII\cite{R16}}     & 9.69\%                 & 76.48\%                 & \multicolumn{1}{r}{9.41\%}  & \multicolumn{1}{r|}{67.11\%} \\
                                     & 14.87\%                & 66.82\%                 & \multicolumn{1}{r}{14.38\%} & \multicolumn{1}{r|}{56.79\%} \\ \hline
    \multirow{2}{*}{TD\cite{b12}}              & 9.90\%                 & 42.67\%                 & 9.82\%                      & 9.24\%                       \\
                                     & 14.94\%                & 24.51\%                 & 14.53\%                     & 0.27\%                       \\ \hline
    \multirow{2}{*}{STD}             & 9.92\%                 & 36.44\%                 & 9.82\%                      & 7.15\%                       \\
                                     & 14.91\%                & 14.62\%                 & 14.68\%                     & 0.24\%                       \\ \hline
    STD+FC                           & 9.92\%                 & 9.24\%                  & 9.93\%                      & 0.53\%                       \\
    (Without $\mathcal{S}$)          & 14.88\%                & 4.12\%                  & 14.95\%                     & 0.11\%                       \\ \hline
    STD+FC                           & 9.88\%                 & 7.61\%                  & 9.61\%                      & 0.11\%                       \\
    ($\mathcal{S}$ Sparsity$=0.01$)  & 14.97\%                & 2.42\%                  & 14.49\%                     & 0.06\%                       \\ \hline
    \end{tabular}
\end{table}

We have four control variables to help us control the effectiveness of our compression: the Tucker rank, \(\mathcal{G}\) sparsity, \(\mathcal{S}\) sparsity and the bit tolerance in the run-length coding. In the following experiments we maintain a relative error of 1\% while doing run-length coding. The table \ref{tab:eMBB} shows the minimum rate loss that our model can achieve within 10\% and 15\% compression after adding RLE, core sparsity, and extra sparsity tensor sequentially to the Tucker decomposition, which can show that each of the parts of the model we added contributes to the compression. We can see from the table that the lowest rate loss of STD is lower than that of TD at the same compression rate, which suggests that the sparsity of core tensor can be further utilized to exploit the potential of tucker decomposition on compression. On top of that, the second stage factor compression method can further achieve a lower rate loss, which illustrates the effectiveness of two-stage compression. Also, the table shows that the rate loss can be further reduced when an additional sparse tensor $\mathcal{S}$ is added. 

Table \ref{tab:para} shows our compression performance with different parameters, showing the role that each compression parameter plays in the RL-CR trade-off. It can be seen that decomposing the sparse tensor \(\mathcal{S}\) has a better RL most of the time, although it increase compression rate. However this approach does not perform so consistently because of the non-convexity of the problem, and even increases the rate loss under specific parameters.

\begin{table}[]
    \centering
    \caption{Rate loss and compression ratio with different tucker rank and sparsity for dataset 1.}\label{tab:para}
    \begin{tabular}{|l|l|llll|}
    \hline
    \multirow{2}{*}{Tucker Rank} & \multirow{2}{*}{$\mathcal{G}$ Sparsity} & \multicolumn{2}{l|}{without $\mathcal{S}$}         & \multicolumn{2}{l|}{$\mathcal{S}$ sparsity=0.01} \\ \cline{3-6} 
                                 &                                & RL      & \multicolumn{1}{l|}{CR} & RL                & CR               \\ \hline \hline
    35x30x2                      & 0.5                            & 10.74\% & 10.74\%                 & 9.67\%            & 11.02\%          \\ \cline{1-2}
    25x30x2                      & 0.5                            & 11.26\% & 9.08\%                  & 10.18\%           & 9.36\%           \\ \cline{1-2}
    30x30x2                      & 0.5                            & 10.66\% & 9.93\%                  & 9.58\%            & 10.21\%          \\ \cline{1-2}
    \multirow{3}{*}{25x40x2}     & 0.3                            & 8.12\%  & 10.05\%                 & 8.26\%            & 10.33\%          \\
                                 & 0.4                            & 6.49\%  & 10.38\%                 & 6.16\%            & 10.66\%          \\
                                 & 0.5                            & 5.43\%  & 10.70\%                 & 4.83\%            & 10.98\%          \\ \cline{1-2}
    \multirow{3}{*}{20x40x2}     & 0.3                            & 10.78\% & 9.22\%                  & 9.56\%            & 9.50\%           \\
                                 & 0.4                            & 9.06\%  & 9.48\%                  & 8.10\%            & 9.76\%           \\
                                 & 0.5                            & 8.17\%  & 9.74\%                  & 7.05\%            & 10.02\%          \\ \cline{1-2}
    \multirow{3}{*}{30x40x2}     & 0.3                            & 6.77\%  & 10.85\%                 & 7.64\%            & 11.13\%          \\
                                 & 0.4                            & 5.66\%  & 11.24\%                 & 5.76\%            & 11.52\%          \\
                                 & 0.5                            & 4.52\%  & 11.63\%                 & 4.42\%            & 11.91\%          \\ \cline{1-2}
    \multirow{3}{*}{25x35x2}     & 0.3                            & 10.50\% & 9.34\%                  & 10.41\%           & 9.62\%           \\
                                 & 0.4                            & 8.87\%  & 9.62\%                  & 8.27\%            & 9.90\%           \\
                                 & 0.5                            & 7.62\%  & 9.90\%                  & 6.92\%            & 10.18\%          \\ \hline
    \end{tabular}
\end{table}

\subsection{Comparison with other methods}
\begin{figure*}[htbp]
    \centering
    \subfloat[dataset 1]{\includegraphics[width=0.45\linewidth]{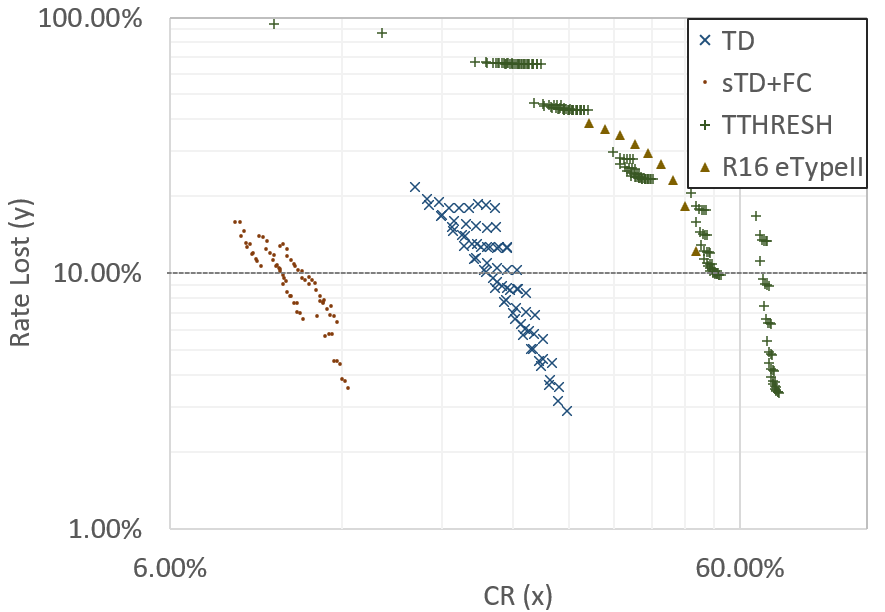}\label{fig:sV}}
    \subfloat[dataset 2]{\includegraphics[width=0.45\linewidth]{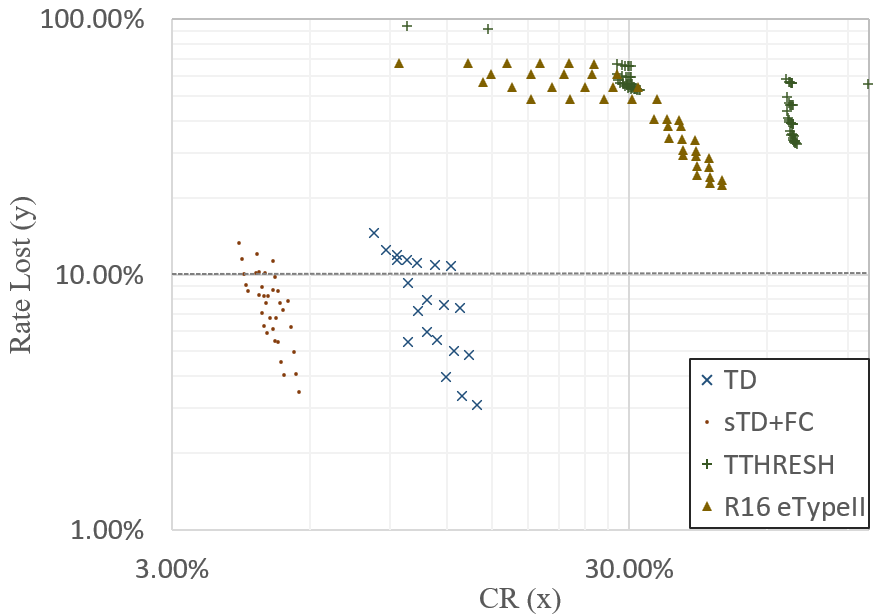}\label{fig:gV}}
    \caption{CR-RL plot of compression of dataset 1 and dataset 2 data by different methods}
\end{figure*}

Since we have two optimization goals: smaller rate loss and lower compression rate, we plot the compression rate versus rate loss for different methods with different parameter choices. In Figure \ref{fig:sV} and \ref{fig:gV}, we report the rate loss (RL) of TD, TTHRESH, STD+FC and R16-eTypeII under different CR for dataset 1 and dataset 2, where STD+FC stands for the fact that on top of STD we also use a compression strategy based on Givens transform and RLE. We can find that TD and STD can reach better compressibility because they utilize the low-rank nature of the weight tensor, i.e., the correlation between the various dimensions of the data. Especially, the compression effect is more obvious on dataset 2, because dataset 2 is larger in scale and has more data redundancy, thus the low-rankness is stronger, and the compressible space is larger. From the figure, it can be seen that our method can achieve a lower compression rate than the other methods for the same rate loss, and table \ref{tab:eMBB} shows that the rate loss of our method is much smaller than the other methods for the same compression rate.

\subsection{Characteristics of compression}
\begin{figure}
    \centering
    \includegraphics[width=\linewidth]{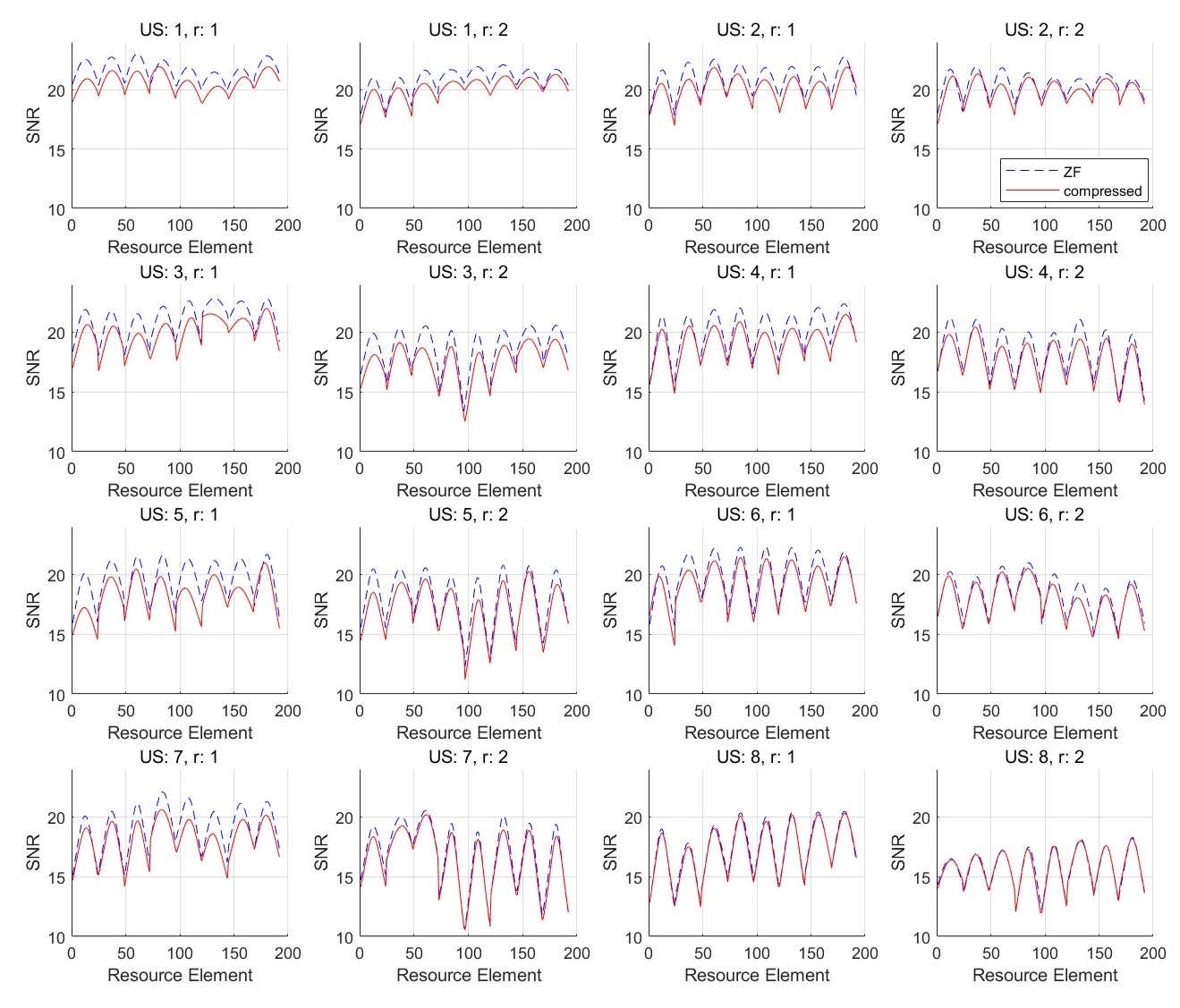}
    \caption{SNR curves under different data streams for different users. The dashed line represents the SNR curve for uncompressed ZF weights, and the solid line represents the SNR curve for compressed ZF weights, as shown in the legend. Here we only show the data of the first 16 RBs.}\label{fig:snr}
\end{figure}

We then evaluate the effect of the compressed weights by their numerical performance on different REs. As Figure \ref{fig:snr} exhibits, the SNR of our compressed weights loses more at the ``spikes'' and holds up better at the troughs, which is most likely caused by the low rank model losing some of its details. Among other things, the compressed weights work as well as the uncompressed ones for user 8 because the associated weights tensor has a very good low rank that STD can fit well.

\section{Conclusion}
This paper inspired by the bottlenecks of downlink eCPRI in the future proposes a two-stage approach for compressing beamforming weight, which involves a tensor sparse tucker decomposition model with run-length encoding for the core tensor and the factor matrices after givens decomposition. We propose our algorithm for this nonconvex model and prove its convergence. By employing this method, we can significantly reduce the downlink eCPRI bandwidth without the need to transmit the original beamforming weights. Numerical experiments can show that we achieve the best compression rate with the smallest rate loss in the existing methods, which verifies that our method is effective. Moreover, our approach has potential applications in various wireless communication scenarios, including massive MIMO systems and wireless sensor networks, where data compression is crucial due to limited communication resources.

However, we still have a lot of room for improvement in this approach. Our choice of parameters for the compression algorithm is based on experience, for which we lack a theoretical guide for choosing the parameters. In future work, we can consider error estimates for low-rank sparse tucker decomposition approximations as well as quantization strategies to guide the corresponding parameter selection. We can also consider dynamically adjusting the algorithm parameters through adaptive methods or other intelligent approaches. At the same time, as illustrated in Fig. \ref{fig:snr}, how to salvage the transmission rate at the ``spikes'' can be a major problem, and solving it can help us to further reduce the rate loss. Furthermore, future research could explore the use of our proposed method in conjunction with other compression techniques to achieve even higher compression rates while minimizing the loss of information. 

\bibliographystyle{IEEEtran.bst}
\bibliography{refer.bib}
\appendix
\subsection{Preliminaries}
To prove the convergence, We first introduce the Kurdyka-Lojasiewicz property, which is used to help us in the analysis of nonconvex problems:

\begin{definition}
     (Kurdyka-Lojasiewicz property) The function $f$ is said to have the KL property at $\bar{x} \in \operatorname{dom} \partial f$ if there exist $\eta \in(0,+\infty]$, a neighborhood $U$ of $\bar{x}$ and a continuous concave function $\varphi:[0, \eta) \rightarrow \mathbb{R}_{+}$such that
\begin{enumerate}
    \item $\varphi(0)=0$;
    \item $\varphi$ is $C^1$ on $(0, \eta)$;
    \item for all $s \in(0, \eta), \varphi^{\prime}(s)>0$;
    \item for all $x$ in $U \cap[F\left(\bar{x}\right)<f<F\left(\bar{x}\right)+\eta]$, the Kurdyka-Lojasiewicz inequality holds, i.e., $$
    \varphi^{\prime}(F\left(x\right)-F\left(\bar{x}\right)) \operatorname{dist}(0, \partial F\left(x\right)) \geq 1 .
    $$
\end{enumerate}
We denote by $\Phi_\eta$ the class of $\varphi$ which satisfies the above definitions (i), (ii) and (iii). If $f$ satisfies the KL property at each point of $\operatorname{dom} \partial f$, then $f$ is called a KL function.
\end{definition}

\begin{lemma} \label{lemma:KL}
 \cite{Bolte2014Proximal}(Uniformized KL property) Let $\Omega$ be a compact set and let $f: \mathbb{R}^n \rightarrow$ $\mathbb{R} \cup\{+\infty\}$ be a proper lower semicontinuous function. Assume that $f$ is constant on $\Omega$ and satisfies the KL property at each point of $\Omega$. Then, there exist $\varepsilon>0, \eta>0$ and $\varphi \in \Phi_\eta$ such that for $\forall \bar{x} \in \Omega$ and all $x$ in the following intersection:
$$
\left\{x \in \mathbb{R}^n: \operatorname{dist}(x, \Omega)<\varepsilon\right\} \cap[F\left(\bar{x}\right)<F\left(x\right)<F\left(\bar{x}\right)+\eta],
$$
one has,
$$
\varphi^{\prime}(F\left(x\right)-F\left(\bar{x}\right)) \operatorname{dist}(0, \partial F\left(x\right)) \geq 1 .
$$
\end{lemma}

Recall that the notation $\mathcal{X}:=(\mathcal{G},\mathbf{U}_1,\mathbf{U}_2,\mathbf{U}_3,\mathcal{S})\in\mathbb{C}^{r_1\times r_2\times r_3}\times\mathbb{C}^{n_1\times r_1}\times\mathbb{C}^{n_2\times r_2}\times\mathbb{C}^{n_3\times r_3}\times\mathbb{C}\times\mathbb{C}^{n_1\times n_2\times n_3}$, we denote that $G(\mathcal{X})=\| \mathcal{V}-\mathcal{S}- [\![\mathcal{G}; \mathbf{U}_1 ,\mathbf{U}_2 ,\mathbf{U}_3 ]\|_F^2$, so $F\left(\mathcal{X}\right)=G(\mathcal{X})+\delta_{\|\mathcal{G}\|_0 \leq \alpha_1}(\mathcal{G})
+\delta_{\|\mathcal{S}\|_0 \leq \alpha_2}(\mathcal{S})+\sum_{i=1}^3\delta_{\mathbf{U}_i^H\mathbf{U}_i=\mathbf{I}}(\mathbf{U}_i).$

For any fixed $\mathbf{U}_1, \mathbf{U}_2, \mathbf{U}_3$ and $\mathcal{S}$, the partial gradient $\nabla_{\mathcal{G}}G(\mathcal{X})$ is globally Lipschitz continuous with
the modulus $l_{\mathcal{G}}(\mathbf{U}_1, \mathbf{U}_2, \mathbf{U}_3, \mathcal{S})$, that is
\begin{align*}
    &\|\nabla_{\mathcal{G}}(G(\mathcal{G}_1,\mathbf{U}_1,\mathbf{U}_2,\mathbf{U}_3,\mathcal{S}))-\nabla_{\mathcal{G}}(\mathcal{G}_2,\mathbf{U}_1,\mathbf{U}_2,\mathbf{U}_3,\mathcal{S})\|\\
    &\leq l_{\mathcal{G}}(\mathbf{U}_1, \mathbf{U}_2, \mathbf{U}_3, \mathcal{S})\|\mathcal{G}_1-\mathcal{G}_2\|,\ \forall\mathcal{G}_1, \mathcal{G}_2\in\mathbb{C}^{r_1\times r_2\times r_3}.
\end{align*}

Under the assumption that the $\left\{\mathcal{X}^k\right\}$ is bounded, for any fixed sequence $\mathbf{U}_1^k, \mathbf{U}_2^k, \mathbf{U}_3^k, \mathcal{S}^k$, there exist $L_{\mathcal{G}}>0$ such that $$\sup\left\{l_{\mathcal{G}}(\mathbf{U}_1^k, \mathbf{U}_2^k, \mathbf{U}_3^k, \mathcal{S}^k,\ k\in\mathbb{N}\right\}\leq L_{\mathcal{G}}. $$ Similarly, we can find $L_1, L_2, L_3$ and $L_\mathcal{S}$ for $\mathbf{U}_1, \mathbf{U}_2, \mathbf{U}_3$ and $L_{\mathcal{S}}$. 

Meanwhile, $G(\mathcal{X})$ is Lipschitz smooth on any bounded subset. In other words, for each bounded $B\subseteq\mathbb{C}^{r_1\times r_2\times r_3}\times\mathbb{C}^{n_1\times r_1}\times\mathbb{C}^{n_2\times r_2}\times\mathbb{C}^{n_3\times r_3}\times\mathbb{C}^{n_1\times n_2\times n_3}$ , there exist $M>0$ such that
\begin{align*}
    &\left\|\left(\begin{aligned}
        \nabla_{\mathcal{G}}G(\mathcal{X}_1)\\
        \nabla_{\mathbf{U}_1}G(\mathcal{X}_1)\\
        \nabla_{\mathbf{U}_2}G(\mathcal{X}_1)\\
        \nabla_{\mathbf{U}_3}G(\mathcal{X}_1)\\
        \nabla_{\mathcal{S}}G(\mathcal{X}_1)\\
    \end{aligned}\right)|_{\mathcal{X}=\mathcal{X}_1}-\left(\begin{aligned}
        \nabla_{\mathcal{G}}G(\mathcal{X}_2)\\
        \nabla_{\mathbf{U}_1}G(\mathcal{X}_2)\\
        \nabla_{\mathbf{U}_2}G(\mathcal{X}_2)\\
        \nabla_{\mathbf{U}_3}G(\mathcal{X}_2)\\
        \nabla_{\mathcal{S}}G(\mathcal{X}_2)\\
    \end{aligned}\right)|_{\mathcal{X}=\mathcal{X}_2}\right\|\leq M\left\|\mathcal{X}_1-\mathcal{X}_2\right\|_F,\ \forall\mathcal{X}_1,\mathcal{X}_2\in B.
\end{align*}

\subsection{Proof of Lamma \ref{lemma:H1}}
\begin{proof}
    (1) To prove the Lemma, we will show the descent property with specific $k$ for each component of $\mathcal{X}$ firstly. For $\mathbf{U}_i$, we have
    \begin{align*}
        &F\left(\mathcal{G}^{k+1},\left\{\mathbf{\bar{U}}^{k+1}_{j<i}\right\},\mathbf{U}_i^{k+1} ,\left\{\mathbf{\bar{U}}^{k}_{j> i}\right\},\mathcal{S}^k\right)
        -F\left(\mathcal{G}^{k+1},\left\{\mathbf{\bar{U}}^{k+1}_{j<i}\right\},\mathbf{U}_i^{k} ,\left\{\mathbf{\bar{U}}^{k}_{j> i}\right\},\mathcal{S}^k\right)
        \\\leq&-\frac{1}{2\eta_i^k}\left\|\mathbf{U}^{k+1}-\mathbf{\bar{U}}^k\right\|_F^2+\frac{1}{2\eta_i^k}\left\|\mathbf{U}^{k}_i-\mathbf{\bar{U}}_i^k\right\|_F^2, 
    \end{align*} 
    which can be rewriten into
    \begin{align*}
        &\delta_{\mathbf{U}_i^H\mathbf{U}_i=\mathbf{I}}(\mathbf{U}^{k+1}_i)+r_i^k(\mathbf{U}_i^{k+1})-\delta_{\mathbf{U}_i^H\mathbf{U}_i=\mathbf{I}}(\mathbf{U}^{k}_i)-r_i^k(\mathbf{U}_i^{k})\\
        \leq &\bar{r}_i^k(\mathbf{U}_i^{k})-\bar{r}_i^k(\mathbf{U}_i^{k+1})-r_i^k(\mathbf{U}_i^{k})+r_i^k(\mathbf{U}_i^{k+1})
        -\frac{1}{2\eta_i^k}\|\mathbf{U}^{k+1}_i-\mathbf{\bar{U}}_i^k\|_F^2+\frac{1}{2\eta_i^k}\|\mathbf{U}^{k}_i-\mathbf{\bar{U}}_i^k\|_F^2,
    \end{align*}
    where $r_i^k(\mathbf{U}_i)=\frac{1}{2}\|\mathcal{V}-\mathcal{S}^k-[\![\mathcal{G}^{k+1};\{\mathbf{U}^{k+1}_{j< i}\}, \mathbf{U}_i ,\{\mathbf{U}^{k}_{j> i}\} ]\!]\|_F^2$ and $\bar{r}_i^k(\mathbf{U}_i)=\frac{1}{2}\|\mathcal{V}-\mathcal{S}^k-[\![\mathcal{G}^{k+1};\{\mathbf{\bar{U}}^{k+1}_{j< i}\}, \mathbf{U}_i ,\{\mathbf{\bar{U}}^{k}_{j> i}\} ]\!]\|_F^2$. Use $\langle x,y\rangle\leq\frac{s}{2}x^2+\frac{1}{2s}y^2$ and the L-smoothness of $\|\cdot\|_F^2$, there exist $M>0$, $L_i>0$ and any $s_i > 0$, such that
    \begin{align}
        \nonumber
        &\bar{r}_i^k(\mathbf{U}_i^{k})-\bar{r}_i^k(\mathbf{U}_i^{k+1})-r_i^k(\mathbf{U}_i^{k})+r_i^k(\mathbf{U}_i^{k+1})
        \nonumber
        -\frac{1}{2\eta_i^k}\|\mathbf{U}^{k+1}_i-\mathbf{\bar{U}}_i^k\|_F^2+\frac{1}{2\eta_i^k}\|\mathbf{U}^{k}_i-\mathbf{\bar{U}}_i^k\|_F^2 
        \\\nonumber
        \leq&\langle\mathbf{U}_i^{k}-\mathbf{U}_i^{k+1},\nabla r_i^k(\mathbf{U}^{k}_i)-\nabla\bar{r}_i^k(\mathbf{U}_i^{k+1})\rangle
        \nonumber
        +L_i\|\mathbf{U}^{k+1}_i-\mathbf{U}_i^k\|_F^2-\frac{1}{2\eta_i^k}\|\mathbf{U}^{k+1}_i-\mathbf{\bar{U}}_i^k\|_F^2
        \nonumber
        +\frac{1}{2\eta_i^k}\|\mathbf{U}^{k}_i-\mathbf{\bar{U}}_i^k\|_F^2 
        \\\nonumber
        \leq&\frac{s_i}{2}\|\nabla r_i^k(\mathbf{U}^{k}_i)-\nabla\bar{r}_i^k(\mathbf{U}_i^{k+1})\|_F^2+\frac{1}{2\eta_i^k}\|\mathbf{U}^{k}_i-\mathbf{\bar{U}}_i^k\|_F^2
        \nonumber
        +(L_i+\frac{1}{2s_i})\|\mathbf{U}^{k+1}_i-\mathbf{U}_i^k\|_F^2-\frac{1}{2\eta_i^k}\|\mathbf{U}^{k+1}_i-\mathbf{\bar{U}}_i^k\|_F^2
        \\\nonumber
        \leq&\frac{M^2s_i}{2}\sum_{j<i}\|\mathbf{U}^{k+1}_j-\mathbf{\bar{U}}^{k+1}_j\|_F^2+\frac{M^2s_i}{2}\sum_{j>i}\|\mathbf{U}^{k}_j-\mathbf{\bar{U}}^{k}_j\|_F^2
        \nonumber
        -\frac{1}{2\eta_i^k}\|\mathbf{U}^{k+1}_i-\mathbf{\bar{U}}_i^k\|_F^2+\frac{1}{2\eta_i^k}\|\mathbf{U}^{k}_i-\mathbf{\bar{U}}_i^k\|_F^2
        \\
        &+(L_i+\frac{1}{2s_i}+\frac{M^2s_i}{2})\|\mathbf{U}^{k+1}_i-\mathbf{U}_i^k\|_F^2.\label{ieq:Ubefore}
    \end{align}
    Notice that (\ref{ieq:Ubefore}) hold for any $s_i$, we set $s_i=\frac{1}{M}$ for simplify, which implies
    \begin{align}
        \nonumber
        &F\left(\mathcal{G}^{k+1},\{\mathbf{U}^{k+1}_{j<i}\},\mathbf{U}_i^{k+1} ,\{\mathbf{U}^{k}_{j> i}\},\mathcal{S}^k\right)
        \nonumber
        -F\left(\mathcal{G}^{k+1},\{\mathbf{U}^{k+1}_{j<i}\},\mathbf{U}_i^{k} ,\{\mathbf{U}^{k}_{j> i}\},\mathcal{S}^k\right), \nonumber
        \\\nonumber
        \leq & \frac{M}{2}\sum_{j<i}\|\mathbf{U}^{k+1}_j-\mathbf{\bar{U}}^{k+1}_j\|_F^2+\frac{M}{2}\sum_{j>i}\|\mathbf{U}^{k}_j-\mathbf{\bar{U}}^{k}_j\|_F^2
        \nonumber
        +(L_i+M)\|\mathbf{U}^{k+1}_i-\mathbf{U}_i^k\|_F^2-\frac{1}{2\eta_i^k}\|\mathbf{U}^{k+1}_i-\mathbf{\bar{U}}_i^k\|_F^2
        \\\label{ieq:UH1}
        &+\frac{1}{2\eta_i^k}\|\mathbf{U}^{k}_i-\mathbf{\bar{U}}_i^k\|_F^2.
    \end{align} 
    
    As for $\mathcal{G}$ and $\mathcal{S}$, we can see that
    \begin{align*}
        \eta_{\mathcal{G}}^k F\left(\mathcal{G}^{k+1},\mathbf{\bar{U}}_1^k,\mathbf{\bar{U}}_2^k,\mathbf{\bar{U}}_3^k,\mathcal{S}^k\right)+\frac{1}{2}\|\mathcal{G}^{k+1}-\mathcal{G}^k\|_F^2
        \leq&\eta_{\mathcal{G}}^k F\left(\mathcal{G}^{k},\mathbf{\bar{U}}_1^k,\mathbf{\bar{U}}_2^k,\mathbf{\bar{U}}_3^k,\mathcal{S}^k\right) 
        \\
        \eta_{\mathcal{S}}^k F\left(\mathcal{G}^{k+1},\mathbf{\bar{U}}_1^{k+1},\mathbf{\bar{U}}_2^{k+1},\mathbf{\bar{U}}_3^{k+1},\mathcal{S}^{k+1}\right)+\frac{1}{2}\|\mathcal{S}^{k+1}-\mathcal{S}^k\|_F^2
        \leq&\eta_{\mathcal{S}}^k F\left(\mathcal{G}^{k+1},\mathbf{\bar{U}}_1^{k+1},\mathbf{\bar{U}}_2^{k+1},\mathbf{\bar{U}}_3^{k+1},\mathcal{S}^{k}\right).
    \end{align*}
    We can similarly obtain the following inequation,
    \begin{equation}\label{ieq:GH1}
        \begin{aligned}
            &F\left(\mathcal{G}^{k+1},\mathbf{U}_1^k,\mathbf{U}_2^k,\mathbf{U}_3^k,\mathcal{S}^k\right)
            -F\left(\mathcal{G}^{k},\mathbf{U}_1^k,\mathbf{U}_2^k,\mathbf{U}_3^k,\mathcal{S}^k\right)\\
            \leq & \frac{M}{2}\sum_{i=1}^3\|\mathbf{U}^k_i-\mathbf{\bar{U}}^k_i\|_F^2
            +(L_{\mathcal{G}}+M-\frac{1}{2\eta_{\mathcal{G}}^k})\|\mathcal{G}^{k+1}-\mathcal{G}^k\|_F^2
        \end{aligned}
    \end{equation}
    
    \begin{equation}\label{ieq:SH1}
        \begin{aligned}
            &F\left(\mathcal{G}^{k+1},\mathbf{U}_1^{k+1},\mathbf{U}_2^{k+1},\mathbf{U}_3^{k+1},\mathcal{S}^{k+1}\right)
            -F\left(\mathcal{G}^{k+1},\mathbf{U}_1^{k+1},\mathbf{U}_2^{k+1},\mathbf{U}_3^{k+1},\mathcal{S}^{k}\right) \\
            \leq & \frac{M}{2}\sum_{i=1}^3\|\mathbf{U}^{k+1}_i-\mathbf{\bar{U}}^{k+1}_i\|_F^2\
            +(L_{\mathcal{S}}+M-\frac{1}{2\eta_{\mathcal{S}}^k})\|\mathcal{S}^{k+1}-\mathcal{S}^k\|_F^2.
        \end{aligned} 
    \end{equation}
    Combine (\ref{ieq:UH1}), (\ref{ieq:GH1}) and (\ref{ieq:SH1}), we will have
    \begin{equation}\label{ieq:H1b1}
        \begin{aligned}
            &F\left(\mathcal{X}^k\right) - F\left(\mathcal{X}^{k+1}\right)
            \\
            \leq & \sum_{i=1}^3(L_i+M)\|\mathbf{U}^{k+1}_i-\mathbf{U}_i^k\|_F^2
            -\frac{1}{2\eta_i^k}\|\mathbf{U}^{k+1}_i-\mathbf{\bar{U}}_i^k\|_F^2 
            +\sum_{i=1}^3(\frac{Mi}{2}+\frac{1}{2\eta_i^k})\|\mathbf{U}^k_i-\mathbf{\bar{U}}^k_i\|_F^2
            \\
            & +(L_{\mathcal{G}}+M-\frac{1}{2\eta_{\mathcal{G}}^k})\|\mathcal{G}^{k+1}-\mathcal{G}^k\|_F^2 
             +\sum_{i=1}^3\frac{M(4-i)}{2}\|\mathbf{U}^{k+1}_i-\mathbf{\bar{U}}^{k+1}_i\|_F^2
            +(L_{\mathcal{S}}+M-\frac{1}{2\eta_{\mathcal{S}}^k})\|\mathcal{S}^{k+1}-\mathcal{S}^k\|_F^2 
        \end{aligned}
    \end{equation}
    By the definition of $\mathbf{\bar{U}}_i$, we can get
    \begin{align*}
        \bar{\mathbf{U}}_i^{k+1}-\mathbf{U}_i^{k+1}&=\beta(\mathbf{U}_i^{k+1}-\bar{\mathbf{U}}_i^{k})\\
        &=\beta(\mathbf{U}_i^{k+1}-\mathbf{U}_i^{k}+\mathbf{U}_i^{k}-\bar{\mathbf{U}}_i^{k})
    \end{align*}
    
    Then we are going to bound $\|\mathbf{U}_i^{k+1}-\mathbf{U}_i^k\|_F^2$ by $\|\mathbf{U}_i^{k+1}-\mathbf{\bar{U}}^{k+1}_i\|_F^2$ and $\|\mathbf{U}_i^{k}-\mathbf{\bar{U}}^k_i\|_F^2$,
    \begin{align}
        \|\mathbf{U}_i^{k+1}-\bar{\mathbf{U}}_i^{k}\|_F^2=&\frac{1}{\beta^2}\|\bar{\mathbf{U}}_i^{k+1}-\mathbf{U}_i^{k+1}\|_F^2 \label{ieq:H1b2}\\
        \|\mathbf{U}_i^{k+1}-\mathbf{U}_i^{k}\|_F^2=&\frac{1}{\beta^2}\|(\mathbf{U}_i^{k+1}-\mathbf{\bar{U}}^{k+1}_i)-\beta(\mathbf{U}_i^{k}-\mathbf{\bar{U}}^k_i)\|_F^2 \nonumber\\
        =&\frac{1}{\beta^2}\|\mathbf{U}_i^{k+1}-\mathbf{\bar{U}}^{k+1}_i\|_F^2+1\|\mathbf{U}_i^{k}-\mathbf{\bar{U}}^k_i\|_F^2
        \nonumber\\
        &-\frac{2}{\beta}\langle\bar{\mathbf{U}}^{k+1}-\mathbf{U}^{k+1}, \bar{\mathbf{U}}^{k}-\mathbf{U}^{k}\rangle \nonumber\\
        \leq&(\frac{1}{\beta^2}+\frac{1}{\beta})\|\mathbf{U}_i^{k+1}-\mathbf{\bar{U}}^{k+1}_i\|_F^2
        \nonumber\\
        &+(1+\frac{1}{\beta})\|\mathbf{U}_i^{k}-\mathbf{\bar{U}}^k_i\|_F^2. \label{ieq:H1b3}
    \end{align}
    
    Combining (\ref{ieq:H1b1}), (\ref{ieq:H1b2}) and (\ref{ieq:H1b3}), we will have
    \begin{equation}\label{ieq:H1b4}
        \begin{aligned}
            &[F\left(\mathcal{X}^{k+1}\right)+\sum_{i=1}^3c^{k}_i\|\bar{\mathbf{U}}_i^{k+1}-\mathbf{U}_i^{k+1}\|_F^2] 
            - [F\left(\mathcal{X}^k\right)+\sum_{i=1}^3b^k_i\|\bar{\mathbf{U}}_i^{k}-\mathbf{U}_i^{k}\|_F^2]
            \\
            \leq &(\frac{1}{2\eta_{\mathcal{S}}^k}-L_{\mathcal{S}}-M)\|\mathcal{S}^{k+1}-\mathcal{S}^k\|_F^2
            +(\frac{1}{2\eta_{\mathcal{G}}^k}-L_{\mathcal{G}}-M)\|\mathcal{G}^{k+1}-\mathcal{G}^k\|_F^2 ,
            \end{aligned}
    \end{equation}
    where $b_i^k=\frac{1}{2\eta_i^k}+\frac{Mi}{2}+\frac{(L_i+M)(1+\beta)}{\beta}$ and $c_i^k=\frac{1}{2\eta_i^k\beta^2}-\frac{M(4-i)}{2}-\frac{(L_i+M)(1+\beta)}{\beta^2}$.

    Let $\epsilon>0$ be real number that satisfies $\beta^2<\frac{1-\epsilon}{1+\epsilon}$, and \begin{align*}
        \eta_i^k&=\eta_i=\frac{(1+\epsilon-\beta^2(1-\epsilon))}{5\beta^2M(1+\epsilon)-2Mi\epsilon\beta^2+2t_i},\\
        \gamma_i&=\frac{1+\beta^2}{4\eta_i\beta^2}+\frac{2Mi-5}{4}+\frac{(L_i+M)(1+\beta)(\beta-1)}{2\beta^2},\\
        t_i&=[1+\epsilon+\beta(1-\epsilon)](L_i+M)(1+\beta),
    \end{align*}
    then we have $b_i^k=(1-\epsilon)\gamma_i$ and $c_i^k=(1+\epsilon)\gamma_i$. Recall that \(H(\mathcal{W})=F\left(\mathcal{X}\right)+\sum_{i=1}^3\gamma_i\|\mathbf{U}_i-\mathbf{\bar{U}}_i\|_F^2\), we can rewrite the (\ref{ieq:H1b4}) in the following form,
    \begin{equation}\label{eq:H1e1}
        \begin{aligned}
            &\rho_1\left[\sum_{i=1}^3\left(\left\|\mathbf{U}_i^k-\bar{\mathbf{U}}_i^k\right\|_F^2+\left\|\mathbf{U}_i^{k+1}-\bar{\mathbf{U}}_i^{k+1}\right\|_F^2\right)\right.\\
            &\left.+\|\mathcal{S}^{k+1}-\mathcal{S}^k\|_F^2+\|\mathcal{G}^{k+1}-\mathcal{G}^k\|_F^2\right] \\
            \leq& H\left(\mathcal{W}^k\right)-H\left(\mathcal{W}^{k+1}\right).
        \end{aligned}
    \end{equation}
    where $\rho_1=\min \left\{\epsilon\gamma_1,\epsilon\gamma_2,\frac{1}{2\eta_{\mathcal{S}}^k}-L_{\mathcal{S}}-M,\frac{1}{2\eta_{\mathcal{G}}^k}-L_{\mathcal{G}}-M\right\}$, which will be positive when $\eta_\mathcal{G}^k\leq\frac{2}{L_\mathcal{G}+M}$ and $\eta_\mathcal{S}^k\leq\frac{2}{L_\mathcal{S}+M}$. 
    
    (2) Since $H$ is proper, the monotonically nonincreasing sequence $H\left(\mathcal{W}^k\right)$ converges to some real number $c$. Summing (\ref{eq:H1e1}) from $k=0$ to some interger $N>0$, we get 
    \begin{equation*}
        \begin{aligned}
            &\sum_{k=0}^{N-1}\left[\sum_{i=1}^3\left(\left\|\mathbf{U}_i^k-\bar{\mathbf{U}}_i^k\right\|_F^2+\left\|\mathbf{U}_i^{k+1}-\bar{\mathbf{U}}_i^{k+1}\right\|_F^2\right)
            +\|\mathcal{S}^{k+1}-\mathcal{S}^k\|_F^2+\|\mathcal{G}^{k+1}-\mathcal{G}^k\|_F^2\right] \\
            \leq &\frac{1}{\rho_1}\left(H\left(\mathcal{W}^0\right)-H\left(\mathcal{W}^{N+1}\right)\right).
        \end{aligned}
    \end{equation*}
    With taking limit as $N\rightarrow+\infty$, we will have
    \begin{align*}
        &\sum_{k=0}^{+\infty}\left[\sum_{i=1}^3\left(\left\|\mathbf{U}_i^k-\bar{\mathbf{U}}_i^k\right\|_F^2+\left\|\mathbf{U}_i^{k+1}-\bar{\mathbf{U}}_i^{k+1}\right\|_F^2\right)
        +\|\mathcal{S}^{k+1}-\mathcal{S}^k\|_F^2+\|\mathcal{G}^{k+1}-\mathcal{G}^k\|_F^2\right]\\
        \leq&\frac{1}{\rho_1}\left(H\left(\mathcal{W}^0\right)-c\right)< +\infty.
    \end{align*}
    Combining (\ref{ieq:H1b3}), we obtain
    \begin{align*}
        &\sum_{k=0}^{+\infty}\left\|\mathcal{X}^{k+1}-\mathcal{X}^k\right\|_F^2\\
        \leq&\sum_{k=0}^{+\infty}\left[\sum_{i=1}^3\left(\left\|\mathbf{U}_i^k-\mathbf{U}_i^{k+1}\right\|_F^2\right)
        +\|\mathcal{S}^{k+1}-\mathcal{S}^k\|_F^2+\|\mathcal{G}^{k+1}-\mathcal{G}^k\|_F^2\right]\\
        <& +\infty.
    \end{align*}
\end{proof}
\subsection{Proof of Lamma \ref{lemma:H2}}
\begin{proof}
    According to \cite[Proposition 10.5]{RockWets98} and \cite[Exercise 10.10]{RockWets98}, the sub-gradient of $A$ can be given as
    $$
    \left\{\begin{aligned}
    \partial_{\mathcal{G}} H(\mathcal{G}, \mathbf{U}_1, \mathbf{U}_2, \mathbf{U}_3, \mathcal{S}, \mathbf{\bar{U}}_1, \mathbf{\bar{U}}_2, \mathbf{\bar{U}}_3)
    =&\partial \delta_{\|\mathcal{G}\|_0 \leq \alpha_1}(\mathcal{G})+\nabla_{\mathcal{G}}H(\mathcal{G},\mathbf{U}_1,\mathbf{U}_2,\mathbf{U}_3,\mathcal{S}), \\
    \partial_{\mathbf{U}_i} H(\mathcal{G}, \mathbf{U}_1, \mathbf{U}_2, \mathbf{U}_3, \mathcal{S}, \mathbf{\bar{U}}_1, \mathbf{\bar{U}}_2, \mathbf{\bar{U}}_3)
    =&\partial \delta_{\mathbf{U}_i^H\mathbf{U}_i=\mathbf{I}}(\mathbf{U}_i)+2 \gamma_i(\mathbf{U}_i-\mathbf{\bar{U}}_i) \\
    &+\nabla_{\mathbf{U}_i}H(\mathcal{G},\mathbf{U}_1,\mathbf{U}_2,\mathbf{U}_3,\mathcal{S}),\\
    \partial_{\mathcal{G}} A(\mathcal{G}, \mathbf{U}_1, \mathbf{U}_2, \mathbf{U}_3, \mathcal{S}, \mathbf{\bar{U}}_1, \mathbf{\bar{U}}_2, \mathbf{\bar{U}}_3)
    =&\partial \delta_{\|\mathcal{G}\|_0 \leq \alpha_1}(\mathcal{S})+\nabla_{\mathcal{S}}H(\mathcal{G},\mathbf{U}_1,\mathbf{U}_2,\mathbf{U}_3,\mathcal{S}), \\
    \partial_{\mathbf{\bar{U}}_i} A(\mathcal{G}, \mathbf{U}_1, \mathbf{U}_2, \mathbf{U}_3, \mathcal{S}, \mathbf{\bar{U}}_1, \mathbf{\bar{U}}_2, \mathbf{\bar{U}}_3)
    =&2 \gamma_i(\mathbf{\bar{U}}_i-\mathbf{U}_i) .
    \end{aligned}\right.
    $$
    By the first order optimality conditions, we have
    \begin{align*}
        0&\in\partial\delta_{\|\mathcal{G}\|_0 \leq \alpha_1}(\mathcal{G}^{k+1})+\frac{1}{\eta_{\mathcal{G}}^k}(\mathcal{G}^{k+1}-\mathcal{G}^k)
        +\nabla_{\mathcal{G}}H(\mathcal{G}^{k+1},\mathbf{\bar{U}}_1^k,\mathbf{\bar{U}}_2^k,\mathbf{\bar{U}}_3^k,\mathcal{S}^{k})\\
        0&\in\partial\delta_{\mathbf{U}_i^H\mathbf{U}_i=\mathbf{I}}(\mathbf{U}^{k+1}_i)+\frac{1}{\eta_{i}^k}(\mathbf{U}^{k+1}-\mathbf{\bar{U}}^k)
        +\nabla_{\mathbf{U}_i}H(\mathcal{G}^{k+1},\{\mathbf{\bar{U}}^{k+1}_{j< i}\}, \mathbf{U}^{k+1}_i ,\{\mathbf{\bar{U}}^{k}_{j> i}\},\mathcal{S}^{k})\\
        0&\in\partial\delta_{\|\mathcal{S}\|_0 \leq \alpha_2}(\mathcal{S}^{k+1})+\frac{1}{\eta_{\mathcal{S}}^k}(\mathcal{S}^{k+1}-\mathcal{S}^k)
        +\nabla_{\mathcal{S}}H(\mathcal{G}^{k+1},\mathbf{\bar{U}}_1^{k+1},\mathbf{\bar{U}}_2^{k+1},\mathbf{\bar{U}}_3^{k+1},\mathcal{S}^{k+1}).
    \end{align*}
    Then we get that $C^{k+1}\in\partial A(\mathcal{W}^{k+1})$. Since $\mathcal{\hat{X}}$ is bounded and $\nabla H$ is Lipschitz continuous on bounded subset, there exist $M>0$ such that
    \begin{align*}
        & \left\|\nabla_{\mathcal{G}}H(\mathcal{G}^{k+1},\mathbf{U}_1^{k+1},\mathbf{U}_2^{k+1},\mathbf{U}_3^{k+1},\mathcal{S}^{k+1})
        -\nabla_{\mathcal{G}}H(\mathcal{G}^{k+1},\mathbf{\bar{U}}_1^k,\mathbf{\bar{U}}_2^k,\mathbf{\bar{U}}_3^k,\mathcal{S}^{k})\right\|_F\\
        \leq&M\left(\sum_{i=1}^3\|\mathbf{\bar{U}}_i^k-\mathbf{U}^{k+1}_i\|_F+\|\mathcal{S}^{k}-\mathcal{S}^{k+1}\|_F\right)
        \\=&M\left(\sum_{i=1}^3\frac{1}{\beta}\|\mathbf{\bar{U}}_i^{k+1}-\mathbf{U}^{k+1}_i\|_F+\|\mathcal{S}^{k}-\mathcal{S}^{k+1}\|_F\right)
    \end{align*}
    \begin{align*}
        & \left\|\nabla_{\mathbf{U}_i}H(\mathcal{G}^{k+1},\mathbf{U}_1^{k+1},\mathbf{U}_2^{k+1},\mathbf{U}_3^{k+1},\mathcal{S}^{k+1})
        -\nabla_{\mathbf{U}_i}H(\mathcal{G}^{k+1},\{\mathbf{\bar{U}}^{k+1}_{j< i}\}, \mathbf{U}^{k+1}_i ,\{\mathbf{\bar{U}}^{k}_{j> i}\},\mathcal{S}^{k})\right\|_F\\
        \leq&M\left(\sum_{j<i}\|\mathbf{\bar{U}}_j^{k+1}-\mathbf{U}^{k+1}_j\|_F
        +\sum_{j>i}\|\mathbf{\bar{U}}_j^{k}-\mathbf{U}^{k+1}_j\|_F+\|\mathcal{S}^{k}-\mathcal{S}^{k+1}\|_F\right)\\
        \leq&M\left(\sum_{j<i}\|\mathbf{\bar{U}}_j^{k+1}-\mathbf{U}^{k+1}_j\|_F
        +\sum_{j>i}\frac{1}{\beta}\|\mathbf{\bar{U}}_j^{k+1}-\mathbf{U}^{k+1}_j\|_F+\|\mathcal{S}^{k}-\mathcal{S}^{k+1}\|_F\right)
    \end{align*}
    \begin{align*}
        & \|\nabla_{\mathcal{S}}H(\mathcal{G}^{k+1},\mathbf{U}_1^{k+1},\mathbf{U}_2^{k+1},\mathbf{U}_3^{k+1},\mathcal{S}^{k+1})-\nabla_{\mathcal{S}}H(\mathcal{G}^{k+1},\mathbf{\bar{U}}_1^{k+1},\mathbf{\bar{U}}_2^{k+1},\mathbf{\bar{U}}_3^{k+1},\mathbf{\bar{U}}_3^{k+1},\mathcal{S}^{k+1})\|_F\\
        \leq&M\left(\sum_{i=1}^3\|\mathbf{\bar{U}}_i^{k+1}-\mathbf{U}^{k+1}_i\|_F\right).
    \end{align*}
    Thus we can get the result with $\rho_2=\min\{\frac{1}{\eta_{\mathcal{G}}},\frac{1}{\eta_{\mathcal{S}}}+4M,\frac{Mi}{\beta}+M(4-i)+4\gamma_i+\frac{1}{\eta_i}|i=1,2,3\}$
    \begin{align*}
        &\|C^{k+1}\|_F\\
        \leq& \|C_{\mathcal{G}}^{k+1}\|_F+\sum_{i=1}^{3}\|C_{1i}^{k+1}\|_F+\|C_{\mathcal{S}}^{k+1}\|_F+\sum_{i=1}^{3}\|C_{2i}^{k+1}\|_F\\
        \leq& \frac{1}{\eta_{\mathcal{G}}}\|\mathcal{G}^{k+1}-\mathcal{G}^k\|_F+\left(\frac{1}{\eta_{\mathcal{S}}}+4M\right)\|\mathcal{S}^{k+1}-\mathcal{S}^k\|_F\\
        &+\sum_{i=1}^3\left(\frac{Mi}{\beta}+M(4-i)+4\gamma_i+\frac{1}{\eta_i}\right)\|\mathbf{\bar{U}}_i^{k+1}-\mathbf{U}^{k+1}_i\|_F\\
        \leq&\|\mathcal{G}^{k+1}-\mathcal{G}^k\|_F+\|\mathcal{S}^{k+1}-\mathcal{S}^k\|_F+\sum_{i=1}^3\|\mathbf{\bar{U}}_i^{k+1}-\mathbf{U}^{k+1}_i\|_F.
    \end{align*}
\end{proof}
\subsection{Proof of Lamma \ref{lemma:crit}}
\begin{proof}
    1. The proof is similar to the first part of Lemma 5 (ii) and (iii) in \cite{Bolte2014Proximal} and hence we omit it here.

    2. Let $\mathcal{W}^c=\left(\mathcal{G}^c, \mathbf{U}_1^c, \mathbf{U}_2^c, \mathbf{U}_3^c,\mathcal{S}^c, \bar{\mathbf{U}}_1^c, \bar{\mathbf{U}}_2^c, \bar{\mathbf{U}}_3^c\right)$ be a limit point of $\{\mathcal{\hat{X}}^{k}\}_{k\in\mathbb{N}}$. This means that there is a subsequence $\mathcal{W}^{k_q}$ such that $\mathcal{W}^{k_q}\rightarrow\mathcal{W}^c$ as $q\rightarrow+\inf$. 

    From the iterative step we have $\mathcal{G}^{k_q}\in\left\{\mathcal{G}|\|\mathcal{G}\|_0 \leq \alpha_1\right\}$. Since $\left\{\mathcal{G}|\|\mathcal{G}\|_0 \leq \alpha_1\right\}$ is close, we obtain that $\mathcal{G}^{c}\in\left\{\mathcal{G}|\|\mathcal{G}\|_0 \leq \alpha_1\right\}$. Likely we can get $\mathbf{U}_i^{c}\in\left\{\mathbf{U}_i|\mathbf{U}_i^H\mathbf{U}_i=\mathbf{I}\right\}$ and $\mathcal{S}^{c}\in\left\{a\mathcal{S}|\|\mathcal{S}\|_0 \leq \alpha_1\right\}$. Together with the Lemma \ref{lemma:H1}, we have
    \begin{align*}
        &\lim_{q\rightarrow+\infty}H\left(\mathcal{W}^{k_q}\right)\\
        =& \lim_{q\rightarrow+\infty}\left\{\frac{1}{2} \| \mathcal{V}-\mathcal{S}^{k_q}- [\![\mathcal{G}^{k_q}; \mathbf{U}^{k_q}_1 ,\mathbf{U}^{k_q}_2 ,\mathbf{U}^{k_q}_3]\!]\|_F^2+\sum_{i=1}^3\gamma_i\|\mathbf{U}^{k_q}_i-\mathbf{\bar{U}}^{k_q}_i\|_F^2\right\}\\
        =&\frac{1}{2} \| \mathcal{V}-\mathcal{S}^{c}- [\![\mathcal{G}^{c}; \mathbf{U}^{c}_1 ,\mathbf{U}^{c}_2 ,\mathbf{U}^{c}_3 ]\!]\|_F^2+\sum_{i=1}^3\gamma_i\|\mathbf{U}^{c}_i-\mathbf{\bar{U}}^{c}_i\|_F^2\\
        =&H\left(\mathcal{W}^{c}\right).
    \end{align*}
    Furthermore, it follows from Lemma \ref{lemma:H1} and Lemma \ref{lemma:H2} that $C^{k}\in\partial H\left(\mathcal{W}^{k}\right)$ and $C^{k}\rightarrow 0$ when $k\rightarrow+\infty$. So we have $0\in\partial H\left(\mathcal{W}^{c}\right)$, which implies $\mathcal{W}^{c}$ is the critical point of $H$. Combining the nonincreasing of $H\left(\mathcal{W}^{k}\right)$, we can get $H$ is constant on $\Omega^*$.

    3. define \[B^{k+1}=(B_{\mathcal{G}}^{k+1}, B_{1}^{k+1}, B_{2}^{k+1}, B_{3}^{k+1}, B_{\mathcal{S}}^{k+1}),\] where \[
    \left\{\begin{aligned}
    B_{\mathcal{G}}^{k+1}=
    &\nabla_{\mathcal{G}}H(\mathcal{G}^{k+1},\mathbf{U}_1^{k+1},\mathbf{U}_2^{k+1},\mathbf{U}_3^{k+1},\mathcal{S}^{k+1})\\
    &-\nabla_{\mathcal{G}}H(\mathcal{G}^{k+1},\mathbf{\bar{U}}_1^k,\mathbf{\bar{U}}_2^k,\mathbf{\bar{U}}_3^k,\mathcal{S}^{k})-\frac{1}{\eta_{\mathcal{G}}^k}(\mathcal{G}^{k+1}-\mathcal{G}^k) \\
    B_{i}^{k+1}=
    & \nabla_{\mathbf{U}_i}H(\mathcal{G}^{k+1},\mathbf{U}_1^{k+1},\mathbf{U}_2^{k+1},\mathbf{U}_3^{k+1},\mathcal{S}^{k+1}) \\
    &-\nabla_{\mathbf{U}_i}H(\mathcal{G}^{k+1},\{\mathbf{\bar{U}}^{k+1}_{j< i}\}, \mathbf{U}^{k+1}_i ,\{\mathbf{\bar{U}}^{k}_{j> i}\},\mathcal{S}^{k})-\frac{1}{\eta_{i}^k}(\mathbf{U}^{k+1}-\mathbf{\bar{U}}^k)\\
    B_{\mathcal{S}}^{k+1}=
    &\nabla_{\mathcal{S}}H(\mathcal{G}^{k+1},\mathbf{U}_1^{k+1},\mathbf{U}_2^{k+1},\mathbf{U}_3^{k+1},\mathcal{S}^{k+1})\\
    &-\nabla_{\mathcal{S}}H(\mathcal{G}^{k+1},\mathbf{\bar{U}}_1^{k+1},\mathbf{\bar{U}}_2^{k+1},\mathbf{\bar{U}}_3^{k+1},\mathcal{S}^{k+1})-\frac{1}{\eta_{\mathcal{S}}^k}(\mathcal{S}^{k+1}-\mathcal{S}^k) .
    \end{aligned}\right.\]
    It is obvious that $B^{k+1}\in\partial F\left(\mathcal{X}^{k+1}\right)$. Similarly, we have
    \begin{align*}
        &\|B^{k+1}\|_F\\
        \leq& \|B_{\mathcal{G}}^{k+1}\|_F+\sum_{i=1}^{4}\|B_{i}^{k+1}\|_F+\|B_{\mathcal{S}}^{k+1}\|_F\\
        \leq& \frac{1}{\eta_{\mathcal{G}}}\|\mathcal{G}^{k+1}-\mathcal{G}^k\|_F+\left(\frac{1}{\eta_{\mathcal{S}}}+4M\right)\|\mathcal{S}^{k+1}-\mathcal{S}^k\|_F\\
        &+\sum_{i=1}^3\left(\frac{Mi}{\beta}+M(4-i)+\frac{1}{\eta_i}\right)\|\mathbf{\bar{U}}_i^{k+1}-\mathbf{U}^{k+1}_i\|_F.
    \end{align*}
    Combining Lemma \ref{lemma:H1}, we have $B^{k_q}\rightarrow 0$ when $q\rightarrow+\infty$, which implies $\mathcal{X}^{*}=\left(\mathcal{G}^*, \mathbf{U}_1^*, \mathbf{U}_2^*, \mathbf{U}_3^*,\mathcal{S}^*\right)$ is the critical point of $F$.
\end{proof}

\subsection{Proof of Theorem \ref{theorem:converge}}
\begin{proof}
    According to Lemma \ref{lemma:H1} and Lemma \ref{lemma:crit}, we get that
$$
\lim _{k \rightarrow \infty} H\left(\mathcal{W}^k\right)=H\left(\mathcal{W}^*\right) \text {. }
$$
We consider two cases.

1. If there exists an integer $k^{\prime}$ for which $H\left(\mathcal{W}^{k^{\prime}}\right)=H\left(\mathcal{W}^*\right)$. Rearranging terms of (\ref{eq:H1e1}) and by Lemma \ref{lemma:H1}, for any $k>k^{\prime}$, we have
$$
\begin{aligned}
    &\rho_1\sum_{i=1}^{3}\left\|\mathbf{U}_i^{k+1}-\bar{\mathbf{U}}_i^{k+1}\right\|_F^2\leq H\left(\mathcal{W}^k\right)-H\left(\mathcal{W}^{k+1}\right) \\
     \leq& H\left(\mathcal{W}^{k^{\prime}}\right)-H\left(\mathcal{W}^*\right)=0,\ i=1,2,3.
\end{aligned}
$$
So, we have $\mathbf{U}_i^{k+1}=\bar{\mathbf{U}}_i^{k+1}$ for $i=1,2,3$ and for any $k>k^{\prime}$. Associated with (\ref{ieq:H1b3}), it follows that $\mathcal{U}^{k+1}=\mathcal{U}^k$ for any $k>k^{\prime}$ holds.

2. Now we assume $H\left(\mathcal{W}^k\right)>H\left(\mathcal{\hat{X}}^*\right)$ for all $k \geq 0$. Since $\lim _{k \rightarrow \infty} H\left(\mathcal{W}^k\right)=H\left(\mathcal{W}^*\right)$, it follows that for any $\eta>0$, there exists a nonnegative integer $k_0$ such that $H\left(\mathcal{W}^k\right)<H\left(\mathcal{W}^*\right)+\eta$ for all $k \geq k_0$. From Lemma \ref{lemma:crit} (1), we know that $\lim _{k \rightarrow+\infty} \operatorname{dist}\left(\mathcal{W}^k, \Omega^*\right)=0$. This means that for any $\varepsilon>0$, there exists a positive integer $k_1$ such that $\operatorname{dist}\left(\mathcal{W}^k, \Omega^*\right)<\varepsilon$ for all $k \geq k_1$. Consequently, for all $k>l:=\max \left\{k_0, k_1\right\}$,
$$
\operatorname{dist}\left(\mathcal{W}^k, \Omega^*\right)<\varepsilon, \quad H\left(\mathcal{W}^*\right)<H\left(\mathcal{W}^k\right)<H\left(\mathcal{W}^*\right)+\eta . 
$$
Since $\Omega^*$ is nonempty and compact set and $S$ is constant on $\Omega^*$, applying Lemma \ref{lemma:KL} with $\Omega=\Omega^*$, we deduce that for any $k>l$,
$$
\varphi^{\prime}\left(H\left(\mathcal{W}^k\right)-H\left(\mathcal{W}^*\right)\right) \operatorname{dist}\left(0, \partial H\left(\mathcal{W}^k\right)\right) \geq 1 .
$$
From Lemma \ref{lemma:H2}, we get that
\[
\begin{aligned}
    &\varphi^{\prime}\left(H\left(\mathcal{W}^k\right)-H\left(\mathcal{W}^*\right)\right)\geq\frac{1}{\rho_2}\left(\|\mathcal{G}^{k+1}-\mathcal{G}^k\|_F+\|\mathcal{S}^{k+1}-\mathcal{S}^k\|_F +\sum_{i=1}^3\|\mathbf{\bar{U}}_i^{k+1}-\mathbf{U}^{k+1}_i\|_F\right)^{-1} .
\end{aligned}
\]
On the other hand, from the concavity of $\varphi$, we get that
$$
\begin{aligned}
    &\varphi\left(H\left(\mathcal{W}^k\right)-H\left(\mathcal{W}^*\right)\right)-\varphi\left(H\left(\mathcal{W}^{k+1}\right)-H\left(\mathcal{W}^*\right)\right) \\
    \geq&\varphi^{\prime}\left(H\left(\mathcal{W}^k\right)-H\left(\mathcal{W}^*\right)\right)\left(H\left(\mathcal{W}^k\right)-H\left(\mathcal{W}^{k+1}\right)\right) .
\end{aligned}
$$
For convenience, we define
$$
\Delta_k=\varphi\left(H\left(\mathcal{W}^k\right)-H\left(\mathcal{W}^*\right)\right)-\varphi\left(H\left(\mathcal{W}^{k+1}\right)-H\left(\mathcal{W}^*\right)\right) .
$$
Combining Lemma \ref{lemma:H1} yields for any $k>l$ that
\begin{align*}
    &\left[\sum_{i=1}^3\left(\left\|\mathbf{U}_i^k-\bar{\mathbf{U}}_i^k\right\|_F^2+\left\|\mathbf{U}_i^{k+1}-\bar{\mathbf{U}}_i^{k+1}\right\|_F^2\right)\right.\\
    &\left.+\|\mathcal{S}^{k+1}-\mathcal{S}^k\|_F^2+\|\mathcal{G}^{k+1}-\mathcal{G}^k\|_F^2\right]\\
    \leq& \frac{\rho_2}{\rho_1} \Delta_k\left(\|\mathcal{G}^{k+1}-\mathcal{G}^k\|_F+\|\mathcal{S}^{k+1}-\mathcal{S}^k\|_F\right.\\
    &\left.+\sum_{i=1}^3\|\mathbf{\bar{U}}_i^{k+1}-\mathbf{U}^{k+1}_i\|_F\right) .
\end{align*}
Using respectively the fact that $(a+b)^2 \leq 2\left(a^2+b^2\right)$ and $2 \sqrt{\xi \eta} \leq \xi+\eta$ for all $\xi, \eta \geq 0$, we infer
\begin{align*}
&\frac{\sqrt{2}}{2}\left[\sum_{i=1}^3\left(\left\|\mathbf{U}_i^k-\bar{\mathbf{U}}_i^k\right\|_F+\left\|\mathbf{U}_i^{k+1}-\bar{\mathbf{U}}_i^{k+1}\right\|_F\right)+\|\mathcal{S}^{k+1}-\mathcal{S}^k\|_F+\|\mathcal{G}^{k+1}-\mathcal{G}^k\|_F\right]\\
\leq&\left[\sum_{i=1}^3\left(\left\|\mathbf{U}_i^k-\bar{\mathbf{U}}_i^k\right\|_F^2+\left\|\mathbf{U}_i^{k+1}-\bar{\mathbf{U}}_i^{k+1}\right\|_F^2\right)+\|\mathcal{S}^{k+1}-\mathcal{S}^k\|_F^2+\|\mathcal{G}^{k+1}-\mathcal{G}^k\|_F^2\right]^{\frac{1}{2}} \\
\leq&\left[\frac{\rho_2}{\rho_1} \Delta_k\left(\|\mathcal{G}^{k+1}-\mathcal{G}^k\|_F+\|\mathcal{S}^{k+1}-\mathcal{S}^k\|_F\right.\left.+\sum_{i=1}^3\|\mathbf{\bar{U}}_i^{k+1}-\mathbf{U}^{k+1}_i\|_F\right)\right]^{\frac{1}{2}} \\
\leq& \frac{1}{2}\|\mathcal{G}^{k+1}-\mathcal{G}^k\|_F+\frac{1}{2}\|\mathcal{S}^{k+1}-\mathcal{S}^k\| _F+\frac{1}{2}\sum_{i=1}^3\|\mathbf{\bar{U}}_i^{k+1}-\mathbf{U}^{k+1}_i\|_F+\frac{\rho_2}{2 \rho_1} \Delta_k.
\end{align*}
Then we can get
\begin{equation}\label{ieq:h3mid}
    \begin{aligned}
        &\left[\sum_{i=1}^3\left(\left\|\mathbf{U}_i^k-\bar{\mathbf{U}}_i^k\right\|_F+\left\|\mathbf{U}_i^{k+1}-\bar{\mathbf{U}}_i^{k+1}\right\|_F\right)+\|\mathcal{S}^{k+1}-\mathcal{S}^k\|_F+\|\mathcal{G}^{k+1}-\mathcal{G}^k\|_F\right]\nonumber\\
        \leq& \frac{\sqrt{2}}{2}\|\mathcal{G}^{k+1}-\mathcal{G}^k\|_F+\frac{\sqrt{2}}{2}\|\mathcal{S}^{k+1}-\mathcal{S}^k\|_F+\frac{\sqrt{2}}{2}\sum_{i=1}^3\|\mathbf{\bar{U}}_i^{k+1}-\mathbf{U}^{k+1}_i\|_F+\frac{\sqrt{2} \rho_2}{2 \rho_1} \Delta_k .
    \end{aligned}
\end{equation}
Since $\varphi \geq 0$, we thus have for any $k>l$ that
$$
\begin{aligned}
&\sum_{i=l+1}^k \Delta_i \\
=&\varphi\left(H\left(\mathcal{W}^{l+1}\right)-H\left(\mathcal{W}^*\right)\right)-\varphi\left(H\left(\mathcal{W}^{k+1}\right)-H\left(\mathcal{W}^*\right)\right) \\
\leq&\varphi\left(H\left(\mathcal{W}^{l+1}\right)-H\left(\mathcal{W}^*\right)\right) .
\end{aligned}
$$
Focus on the defination of $\mathbf{\hat{U_i}}$, we can find that
\begin{align*}
    \|\mathbf{U}_i^{k+1}-\mathbf{U}_i^k\|_F=&\frac{1}{\beta}\|(\mathbf{U}_i^{k+1}-\mathbf{\bar{U}}^{k+1}_i)-\beta(\mathbf{U}_i^{k}-\mathbf{\bar{U}}^k_i)\|_F\\
    \leq&\frac{1}{\beta}\|\mathbf{U}_i^{k+1}-\mathbf{\bar{U}}^{k+1}_i\|_F+\|\mathbf{U}_i^{k}-\mathbf{\bar{U}}^k_i\|_F\\
    \leq&m\left[\left(1-\frac{\sqrt{2}}{2}\right)\|\mathbf{U}_i^{k+1}-\mathbf{\bar{U}}^{k+1}_i\|_F+\|\mathbf{U}_i^{k}-\mathbf{\bar{U}}^k_i\|_F\right],
\end{align*}
where $m=\max\{1,\frac{2}{\beta(2-\sqrt{2})}\}$.Then from (\ref{ieq:h3mid}),
\begin{align*}
&\sum_{i=l+1}^k\left[\frac{1}{m}\sum_{i=1}^3\left\|\mathbf{U}_i^k-\mathbf{U}_i^{k+1}\right\|_F+\left(1-\frac{\sqrt{2}}{2}\right)\left(\|\mathcal{S}^{k+1}-\mathcal{S}^k\|_F+\|\mathcal{G}^{k+1}-\mathcal{G}^k\|_F\right)\right]\\
\leq&\sum_{i=l+1}^k\left[\sum_{i=1}^3\left\|\mathbf{U}_i^k-\bar{\mathbf{U}}_i^k\right\|_F+\left(1-\frac{\sqrt{2}}{2}\right)\left(\|\mathcal{S}^{k+1}-\mathcal{S}^k\|_F\right.\left.+\sum_{i=1}^3\left\|\mathbf{U}_i^{k+1}-\bar{\mathbf{U}}_i^{k+1}\right\|_F+\|\mathcal{G}^{k+1}-\mathcal{G}^k\|_F\right)\right]\\
\leq& \frac{\sqrt{2} \rho_2}{2 \rho_1} \varphi\left(H\left(\mathcal{W}^{l+1}\right)-H\left(\mathcal{W}^*\right)\right) .
\end{align*}
This shows the sequence $\left\{\mathcal{X}^k\right\}_{k \in \mathbb{N}}$ is a Cauchy sequence. Indeed, with $q>p>l$, we have
$$
\left\|\mathcal{X}^q-\mathcal{X}^p\right\|=\left\|\sum_{k=p}^{q-1}\left(\mathcal{X}^{k+1}-\mathcal{X}^k\right)\right\| \leq \sum_{k=p}^{q-1}\left\|\mathcal{X}^{k+1}-\mathcal{X}^k\right\| .
$$
This implies $$
\sum_{k=0}^{+\infty}\left\|\mathcal{X}^k-\mathcal{X}^{k+1}\right\|<+\infty.
$$ %TODO: converges to critical of F
\end{proof}
\end{document}